\documentclass[conference,letterpaper]{IEEEtran}
\IEEEoverridecommandlockouts
\usepackage[utf8]{inputenc} 
\usepackage[T1]{fontenc}
\usepackage{url}              
\usepackage{cite}             

\usepackage[cmex10]{amsmath}  
\usepackage{mleftright}       
\mleftright                   

\usepackage{graphicx}         
\usepackage{booktabs}         
\usepackage{multirow}
\usepackage[dvipsnames]{xcolor}
\usepackage{diagbox}

\usepackage[caption=false,font=footnotesize]{subfig}

\usepackage{stfloats}
\usepackage{array}

\usepackage[utf8]{inputenc}
\usepackage{relsize}
\usepackage{amsfonts, amssymb, cuted}
\usepackage{amsthm}
\usepackage{cleveref}
\usepackage{mathtools}
\usepackage{algorithm}
\usepackage[noend]{algpseudocode}
\usepackage{xcolor}
\usepackage{soul}
\usepackage{nicefrac}
\usepackage{enumitem}
\usepackage{acro}
\usepackage{pifont}
\usepackage{tikz}
\usepackage{pgfplotstable}
\usepackage{pgfplots}
\usetikzlibrary{plotmarks}
  \usetikzlibrary{arrows.meta}
  \usepgfplotslibrary{patchplots}
  \usepackage{grffile}
  
  \pgfplotsset{plot coordinates/math parser=false}
  \newlength\figureheight
  \newlength\figurewidth
   \pgfplotsset{compat=1.11,
    /pgfplots/ybar legend/.style={
    /pgfplots/legend image code/.code={%
       \draw[##1,/tikz/.cd,yshift=-0.25em]
        (0cm,0cm) rectangle (3em,8pt);},
   },
}

\title{ Coded Multi-User Information Retrieval\\with a Multi-Antenna Helper Node
}
\author{
\IEEEauthorblockN{Milad Abolpour, MohammadJavad Salehi, Soheil Mohajer, Seyed Pooya Shariatpanahi, and Antti T\"olli} \\
\IEEEauthorblockA{
}
}

 \author{%
   \IEEEauthorblockN{Milad Abolpour\IEEEauthorrefmark{1},
                     MohammadJavad Salehi\IEEEauthorrefmark{1},
                     Soheil Mohajer\IEEEauthorrefmark{2},
                     Seyed Pooya Shariatpanahi\IEEEauthorrefmark{3},
                     and Antti T\"olli\IEEEauthorrefmark{1}}
   \IEEEauthorblockA{\IEEEauthorrefmark{1}%
                     Centre for Wireless Communications, University of Oulu,  Finland, 
                     \textrm{E-mail: \{firstname.lastname\}@oulu.fi}}
   \IEEEauthorblockA{\IEEEauthorrefmark{2}%
                     Department of Electrical and Computer Engineering, University of Minnesota, USA, 
                     \textrm{E-mail: soheil@umn.edu}}
   \IEEEauthorblockA{\IEEEauthorrefmark{3}%
                     School of Electrical and Computer Engineering, 
                     University of Tehran, Iran,
                     \textrm{E-mail: p.shariatpanahi@ut.ac.ir}
                     }
\thanks{
This research has been supported by the Academy of Finland, 6G Flagship program under Grants 346208, 343586 (CAMAIDE), and by the Finnish-American Research and Innovation Accelerator (FARIA).}
\vspace{-9pt}
 }

\begin{document}

\newtheorem{lemma}{Lemma}
\newtheorem{corollary}{Corollary}

\newcommand{\diag}{{\mbox{diag}}}
\newcommand{\herm}{^{\mbox{\scriptsize H}}}
\newcommand{\sherm}{^{\mbox{\scriptsize H}}}
\newcommand{\tran}{^{\mbox{\scriptsize T}}}
\newcommand{\stran}{^{\mbox{\tiny T}}}

\newcommand{\vbar}{\raisebox{.17ex}{\rule{.04em}{1.35ex}}}
\newcommand{\vbarind}{\raisebox{.01ex}{\rule{.04em}{1.1ex}}}
\newcommand{\R}{\ifmmode{\rm I}\hspace{-.2em}{\rm R} \else ${\rm I}\hspace{-.2em}{\rm R}$ \fi}
\newcommand{\T}{\ifmmode{\rm I}\hspace{-.2em}{\rm T} \else ${\rm I}\hspace{-.2em}{\rm T}$ \fi}
\newcommand{\N}{\ifmmode{\rm I}\hspace{-.2em}{\rm N} \else \mbox{${\rm I}\hspace{-.2em}{\rm N}$} \fi}
\newcommand{\B}{\ifmmode{\rm I}\hspace{-.2em}{\rm B} \else \mbox{${\rm I}\hspace{-.2em}{\rm B}$} \fi}
\newcommand{\Hil}{\ifmmode{\rm I}\hspace{-.2em}{\rm H} \else \mbox{${\rm I}\hspace{-.2em}{\rm H}$} \fi}
\newcommand{\C}{\ifmmode\hspace{.2em}\vbar\hspace{-.31em}{\rm C} \else \mbox{$\hspace{.2em}\vbar\hspace{-.31em}{\rm C}$} \fi}
\newcommand{\Cind}{\ifmmode\hspace{.2em}\vbarind\hspace{-.25em}{\rm C} \else \mbox{$\hspace{.2em}\vbarind\hspace{-.25em}{\rm C}$} \fi}
\newcommand{\Q}{\ifmmode\hspace{.2em}\vbar\hspace{-.31em}{\rm Q} \else \mbox{$\hspace{.2em}\vbar\hspace{-.31em}{\rm Q}$} \fi}
\newcommand{\Z}{\ifmmode{\rm Z}\hspace{-.28em}{\rm Z} \else ${\rm Z}\hspace{-.28em}{\rm Z}$ \fi}

\newcommand{\sgn}{\mbox{sgn}}
\newcommand{\var}{\mbox{var}}
\renewcommand{\Re}{\mbox{Re}}
\renewcommand{\Im}{\mbox{Im}}

\renewcommand{\vec}[1]{\bf{#1}}     
\newcommand{\vecsc}[1]{\mbox{\bf \scriptsize #1}}
\newcommand{\itvec}[1]{\mbox{\boldmath{$#1$}}}
\newcommand{\itvecsc}[1]{\mbox{\boldmath{$\scriptstyle #1$}}}
\newcommand{\gvec}[1]{\mbox{\boldmath{$#1$}}}

\newcommand{\balpha}{\mbox{\boldmath{$\alpha$}}}
\newcommand{\bbeta}{\mbox{\boldmath{$\beta$}}}
\newcommand{\bgamma}{\mbox{\boldmath{$\gamma$}}}
\newcommand{\bdelta}{\mbox{\boldmath{$\delta$}}}
\newcommand{\bepsilon}{\mbox{\boldmath{$\epsilon$}}}
\newcommand{\bvarepsilon}{\mbox{\boldmath{$\varepsilon$}}}
\newcommand{\bzeta}{\mbox{\boldmath{$\zeta$}}}
\newcommand{\boldeta}{\mbox{\boldmath{$\eta$}}}
\newcommand{\btheta}{\mbox{\boldmath{$\theta$}}}
\newcommand{\bvartheta}{\mbox{\boldmath{$\vartheta$}}}
\newcommand{\biota}{\mbox{\boldmath{$\iota$}}}
\newcommand{\blambda}{\mbox{\boldmath{$\lambda$}}}
\newcommand{\bmu}{\mbox{\boldmath{$\mu$}}}
\newcommand{\bnu}{\mbox{\boldmath{$\nu$}}}
\newcommand{\bxi}{\mbox{\boldmath{$\xi$}}}
\newcommand{\bpi}{\mbox{\boldmath{$\pi$}}}
\newcommand{\bvarpi}{\mbox{\boldmath{$\varpi$}}}
\newcommand{\brho}{\mbox{\boldmath{$\rho$}}}
\newcommand{\bvarrho}{\mbox{\boldmath{$\varrho$}}}
\newcommand{\bsigma}{\mbox{\boldmath{$\sigma$}}}
\newcommand{\bvarsigma}{\mbox{\boldmath{$\varsigma$}}}
\newcommand{\btau}{\mbox{\boldmath{$\tau$}}}
\newcommand{\bupsilon}{\mbox{\boldmath{$\upsilon$}}}
\newcommand{\bphi}{\mbox{\boldmath{$\phi$}}}
\newcommand{\bvarphi}{\mbox{\boldmath{$\varphi$}}}
\newcommand{\bchi}{\mbox{\boldmath{$\chi$}}}
\newcommand{\bpsi}{\mbox{\boldmath{$\psi$}}}
\newcommand{\bomega}{\mbox{\boldmath{$\omega$}}}

\newcommand{\bolda}{\mbox{\boldmath{$a$}}}
\newcommand{\bb}{\mbox{\boldmath{$b$}}}
\newcommand{\bc}{\mbox{\boldmath{$c$}}}
\newcommand{\bd}{\mbox{\boldmath{$d$}}}
\newcommand{\bolde}{\mbox{\boldmath{$e$}}}
\newcommand{\boldf}{\mbox{\boldmath{$f$}}}
\newcommand{\bg}{\mbox{\boldmath{$g$}}}
\newcommand{\bh}{\mbox{\boldmath{$h$}}}
\newcommand{\bp}{\mbox{\boldmath{$p$}}}
\newcommand{\bq}{\mbox{\boldmath{$q$}}}
\newcommand{\br}{\mbox{\boldmath{$r$}}}
\newcommand{\bs}{\mbox{\boldmath{$s$}}}
\newcommand{\bt}{\mbox{\boldmath{$t$}}}
\newcommand{\bu}{\mbox{\boldmath{$u$}}}
\newcommand{\bv}{\mbox{\boldmath{$v$}}}
\newcommand{\bw}{\mbox{\boldmath{$w$}}}
\newcommand{\bx}{\mbox{\boldmath{$x$}}}
\newcommand{\by}{\mbox{\boldmath{$y$}}}
\newcommand{\bz}{\mbox{\boldmath{$z$}}}

\newtheorem{prop}{Proposition}
\newtheorem{cor}{Corollary}
\newtheorem{conj}{Conjecture}[section]
\newtheorem{exmp}{Example}
\theoremstyle{definition}
\newtheorem{defn}{Definition}
\newtheorem{thm}{Theorem}
\newtheorem{lem}{Lemma}
\newtheorem{rem}{Remark}

\newcommand{\CA}[0]{{\mathcal{A}}}
\newcommand{\CB}[0]{{\mathcal{B}}}
\newcommand{\CC}[0]{{\mathcal{C}}}
\newcommand{\CD}[0]{{\mathcal{D}}}
\newcommand{\CE}[0]{{\mathcal{E}}}
\newcommand{\CF}[0]{{\mathcal{F}}}
\newcommand{\CG}[0]{{\mathcal{G}}}
\newcommand{\CH}[0]{{\mathcal{H}}}
\newcommand{\CI}[0]{{\mathcal{I}}}
\newcommand{\CJ}[0]{{\mathcal{J}}}
\newcommand{\CK}[0]{{\mathcal{K}}}
\newcommand{\CL}[0]{{\mathcal{L}}}
\newcommand{\CM}[0]{{\mathcal{M}}}
\newcommand{\CN}[0]{{\mathcal{N}}}
\newcommand{\CO}[0]{{\mathcal{O}}}
\newcommand{\CP}[0]{{\mathcal{P}}}
\newcommand{\CQ}[0]{{\mathcal{Q}}}
\newcommand{\CR}[0]{{\mathcal{R}}}
\newcommand{\CS}[0]{{\mathcal{S}}}
\newcommand{\CT}[0]{{\mathcal{T}}}
\newcommand{\CU}[0]{{\mathcal{U}}}
\newcommand{\CV}[0]{{\mathcal{V}}}
\newcommand{\CW}[0]{{\mathcal{W}}}
\newcommand{\CX}[0]{{\mathcal{X}}}
\newcommand{\CY}[0]{{\mathcal{Y}}}
\newcommand{\CZ}[0]{{\mathcal{Z}}}

\newcommand{\Ba}[0]{{\mathbf{a}}}
\newcommand{\Bb}[0]{{\mathbf{b}}}
\newcommand{\Bc}[0]{{\mathbf{c}}}
\newcommand{\Bd}[0]{{\mathbf{d}}}
\newcommand{\Be}[0]{{\mathbf{e}}}
\newcommand{\Bf}[0]{{\mathbf{f}}}
\newcommand{\Bg}[0]{{\mathbf{g}}}
\newcommand{\Bh}[0]{{\mathbf{h}}}
\newcommand{\Bi}[0]{{\mathbf{i}}}
\newcommand{\Bj}[0]{{\mathbf{j}}}
\newcommand{\Bk}[0]{{\mathbf{k}}}
\newcommand{\Bl}[0]{{\mathbf{l}}}
\newcommand{\Bm}[0]{{\mathbf{m}}}
\newcommand{\Bn}[0]{{\mathbf{n}}}
\newcommand{\Bo}[0]{{\mathbf{o}}}
\newcommand{\Bp}[0]{{\mathbf{p}}}
\newcommand{\Bq}[0]{{\mathbf{q}}}
\newcommand{\Br}[0]{{\mathbf{r}}}
\newcommand{\Bs}[0]{{\mathbf{s}}}
\newcommand{\Bt}[0]{{\mathbf{t}}}
\newcommand{\Bu}[0]{{\mathbf{u}}}
\newcommand{\Bv}[0]{{\mathbf{v}}}
\newcommand{\Bw}[0]{{\mathbf{w}}}
\newcommand{\Bx}[0]{{\mathbf{x}}}
\newcommand{\By}[0]{{\mathbf{y}}}
\newcommand{\Bz}[0]{{\mathbf{z}}}

\newcommand{\BA}[0]{{\mathbf{A}}}
\newcommand{\BB}[0]{{\mathbf{B}}}
\newcommand{\BC}[0]{{\mathbf{C}}}
\newcommand{\BD}[0]{{\mathbf{D}}}
\newcommand{\BE}[0]{{\mathbf{E}}}
\newcommand{\BF}[0]{{\mathbf{F}}}
\newcommand{\BG}[0]{{\mathbf{G}}}
\newcommand{\BH}[0]{{\mathbf{H}}}
\newcommand{\BI}[0]{{\mathbf{I}}}
\newcommand{\BJ}[0]{{\mathbf{J}}}
\newcommand{\BK}[0]{{\mathbf{K}}}
\newcommand{\BL}[0]{{\mathbf{L}}}
\newcommand{\BM}[0]{{\mathbf{M}}}
\newcommand{\BN}[0]{{\mathbf{N}}}
\newcommand{\BO}[0]{{\mathbf{O}}}
\newcommand{\BP}[0]{{\mathbf{P}}}
\newcommand{\BQ}[0]{{\mathbf{Q}}}
\newcommand{\BR}[0]{{\mathbf{R}}}
\newcommand{\BS}[0]{{\mathbf{S}}}
\newcommand{\BT}[0]{{\mathbf{T}}}
\newcommand{\BU}[0]{{\mathbf{U}}}
\newcommand{\BV}[0]{{\mathbf{V}}}
\newcommand{\BW}[0]{{\mathbf{W}}}
\newcommand{\BX}[0]{{\mathbf{X}}}
\newcommand{\BY}[0]{{\mathbf{Y}}}
\newcommand{\BZ}[0]{{\mathbf{Z}}}

\newcommand{\Bra}[0]{{\Bar{a}}}
\newcommand{\Brb}[0]{{\Bar{b}}}
\newcommand{\Brc}[0]{{\Bar{c}}}
\newcommand{\Brd}[0]{{\Bar{d}}}
\newcommand{\Bre}[0]{{\Bar{e}}}
\newcommand{\Brf}[0]{{\Bar{f}}}
\newcommand{\Brg}[0]{{\Bar{g}}}
\newcommand{\Brh}[0]{{\Bar{h}}}
\newcommand{\Bri}[0]{{\Bar{i}}}
\newcommand{\Brj}[0]{{\Bar{j}}}
\newcommand{\Brk}[0]{{\Bar{k}}}
\newcommand{\Brl}[0]{{\Bar{l}}}
\newcommand{\Brm}[0]{{\Bar{m}}}
\newcommand{\Brn}[0]{{\Bar{n}}}
\newcommand{\Bro}[0]{{\Bar{o}}}
\newcommand{\Brp}[0]{{\Bar{p}}}
\newcommand{\Brq}[0]{{\Bar{q}}}
\newcommand{\Brr}[0]{{\Bar{r}}}
\newcommand{\Brs}[0]{{\Bar{s}}}
\newcommand{\Brt}[0]{{\Bar{t}}}
\newcommand{\Bru}[0]{{\Bar{u}}}
\newcommand{\Brv}[0]{{\Bar{v}}}
\newcommand{\Brw}[0]{{\Bar{w}}}
\newcommand{\Brx}[0]{{\Bar{x}}}
\newcommand{\Bry}[0]{{\Bar{y}}}
\newcommand{\Brz}[0]{{\Bar{z}}}

\newcommand{\BrA}[0]{{\Bar{A}}}
\newcommand{\BrB}[0]{{\Bar{B}}}
\newcommand{\BrC}[0]{{\Bar{C}}}
\newcommand{\BrD}[0]{{\Bar{D}}}
\newcommand{\BrE}[0]{{\Bar{E}}}
\newcommand{\BrF}[0]{{\Bar{F}}}
\newcommand{\BrG}[0]{{\Bar{G}}}
\newcommand{\BrH}[0]{{\Bar{H}}}
\newcommand{\BrI}[0]{{\Bar{I}}}
\newcommand{\BrJ}[0]{{\Bar{J}}}
\newcommand{\BrK}[0]{{\Bar{K}}}
\newcommand{\BrL}[0]{{\Bar{L}}}
\newcommand{\BrM}[0]{{\Bar{M}}}
\newcommand{\BrN}[0]{{\Bar{N}}}
\newcommand{\BrO}[0]{{\Bar{O}}}
\newcommand{\BrP}[0]{{\Bar{P}}}
\newcommand{\BrQ}[0]{{\Bar{Q}}}
\newcommand{\BrR}[0]{{\Bar{R}}}
\newcommand{\BrS}[0]{{\Bar{S}}}
\newcommand{\BrT}[0]{{\Bar{T}}}
\newcommand{\BrU}[0]{{\Bar{U}}}
\newcommand{\BrV}[0]{{\Bar{V}}}
\newcommand{\BrW}[0]{{\Bar{W}}}
\newcommand{\BrX}[0]{{\Bar{X}}}
\newcommand{\BrY}[0]{{\Bar{Y}}}
\newcommand{\BrZ}[0]{{\Bar{Z}}}

\newcommand{\Sfa}[0]{{\mathsf{a}}}
\newcommand{\Sfb}[0]{{\mathsf{b}}}
\newcommand{\Sfc}[0]{{\mathsf{c}}}
\newcommand{\Sfd}[0]{{\mathsf{d}}}
\newcommand{\Sfe}[0]{{\mathsf{e}}}
\newcommand{\Sff}[0]{{\mathsf{f}}}
\newcommand{\Sfg}[0]{{\mathsf{g}}}
\newcommand{\Sfh}[0]{{\mathsf{h}}}
\newcommand{\Sfi}[0]{{\mathsf{i}}}
\newcommand{\Sfj}[0]{{\mathsf{j}}}
\newcommand{\Sfk}[0]{{\mathsf{k}}}
\newcommand{\Sfl}[0]{{\mathsf{l}}}
\newcommand{\Sfm}[0]{{\mathsf{m}}}
\newcommand{\Sfn}[0]{{\mathsf{n}}}
\newcommand{\Sfo}[0]{{\mathsf{o}}}
\newcommand{\Sfp}[0]{{\mathsf{p}}}
\newcommand{\Sfq}[0]{{\mathsf{q}}}
\newcommand{\Sfr}[0]{{\mathsf{r}}}
\newcommand{\Sfs}[0]{{\mathsf{s}}}
\newcommand{\Sft}[0]{{\mathsf{t}}}
\newcommand{\Sfu}[0]{{\mathsf{u}}}
\newcommand{\Sfv}[0]{{\mathsf{v}}}
\newcommand{\Sfw}[0]{{\mathsf{w}}}
\newcommand{\Sfx}[0]{{\mathsf{x}}}
\newcommand{\Sfy}[0]{{\mathsf{y}}}
\newcommand{\Sfz}[0]{{\mathsf{z}}}

\newcommand{\SfA}[0]{{\mathsf{A}}}
\newcommand{\SfB}[0]{{\mathsf{B}}}
\newcommand{\SfC}[0]{{\mathsf{C}}}
\newcommand{\SfD}[0]{{\mathsf{D}}}
\newcommand{\SfE}[0]{{\mathsf{E}}}
\newcommand{\SfF}[0]{{\mathsf{F}}}
\newcommand{\SfG}[0]{{\mathsf{G}}}
\newcommand{\SfH}[0]{{\mathsf{H}}}
\newcommand{\SfI}[0]{{\mathsf{I}}}
\newcommand{\SfJ}[0]{{\mathsf{J}}}
\newcommand{\SfK}[0]{{\mathsf{K}}}
\newcommand{\SfL}[0]{{\mathsf{L}}}
\newcommand{\SfM}[0]{{\mathsf{M}}}
\newcommand{\SfN}[0]{{\mathsf{N}}}
\newcommand{\SfO}[0]{{\mathsf{O}}}
\newcommand{\SfP}[0]{{\mathsf{P}}}
\newcommand{\SfQ}[0]{{\mathsf{Q}}}
\newcommand{\SfR}[0]{{\mathsf{R}}}
\newcommand{\SfS}[0]{{\mathsf{S}}}
\newcommand{\SfT}[0]{{\mathsf{T}}}
\newcommand{\SfU}[0]{{\mathsf{U}}}
\newcommand{\SfV}[0]{{\mathsf{V}}}
\newcommand{\SfW}[0]{{\mathsf{W}}}
\newcommand{\SfX}[0]{{\mathsf{X}}}
\newcommand{\SfY}[0]{{\mathsf{Y}}}
\newcommand{\SfZ}[0]{{\mathsf{Z}}}


\renewcommand{\Re}{\mbox{Re}}
\renewcommand{\Im}{\mbox{Im}}

\renewcommand{\vec}[1]{\bf{#1}}     

\newcommand{\FillGray}[3]{\filldraw[gray!50](#3-1+0.1,#1-#2+0.1) rectangle (#3-0.1,#1-#2+1-0.1)}
\newcommand{\FillBlack}[3]{\filldraw[black!70](#3-1+0.1,#1-#2+0.1) rectangle (#3-0.1,#1-#2+1-0.1)}
\newcommand{\FillHatch}[3]{\fill[pattern=crosshatch, pattern color=black!65](#3-1,#1-#2)rectangle(#3,#1-#2+1)}
\ExplSyntaxOn

\NewDocumentCommand \vect { s o m }
 {
  \IfBooleanTF {#1}
   { \vectaux*{#3} }
   { \IfValueTF {#2} { \vectaux[#2]{#3} } { \vectaux{#3} } }
 }
\DeclarePairedDelimiterX \vectaux [1] {\lbrack} {\rbrack}
 { \, \dbacc_vect:n { #1 } \, }
\cs_new_protected:Npn \dbacc_vect:n #1
 {
  \seq_set_split:Nnn \l_tmpa_seq { , } { #1 }
  \seq_use:Nn \l_tmpa_seq { \enspace }
 }
\ExplSyntaxOff
\maketitle

\begin{abstract}
A novel coding design is proposed to enhance information retrieval in a wireless network of users with partial access to the data, in the sense of observation, measurement, computation, or storage. Information exchange in the network is assisted by a multi-antenna base station (BS), with no direct access to the data. Accordingly, the missing parts of data are exchanged among users through an uplink (UL) step followed by a downlink (DL) step. In this paper, new coding strategies, inspired by coded caching (CC) techniques, are devised to enhance both UL and DL steps. In the UL step, users transmit encoded and properly combined parts of their accessible data to the BS. Then, during the DL step, the BS carries out the required processing on its received signals and forwards a proper combination of the resulting signal terms back to the users, enabling each user to retrieve the desired information. Using the devised coded data retrieval strategy, the data exchange in both UL and DL steps requires the same communication delay, measured by normalized delivery time (NDT). Furthermore, the NDT of the UL/DL step is shown to coincide with the optimal NDT of the original DL multi-input single-output CC scheme, in which the BS is connected to a centralized data library.
\end{abstract}

\section{Introduction}
Multi-user information retrieval (MIR) is a generic field of research exploring mechanisms to enable each user in the network to recover specific pieces of information that are either aggregated at a central master node or distributed across the network~\cite{9463425}. 
An example use case is a distributed coded computing platform where the computation tasks are split among multiple servers, and the outputs are gathered and distributed by a master node (acting as the BS) such that each server has the result of a specific task~\cite{8051074}. Another example is a sensor network where the data is gathered by a BS from sensing nodes and then distributed to multiple actuators, each needing a specific type of data for their action~\cite{XU2004945}. Clearly, with the involvement of a BS, MIR consists of two consecutive steps: 1) an uplink (UL) step where the data is gathered by the BS, and 2) a downlink (DL) step where the gathered data is distributed to requesting nodes according to their needs. The goal of this paper is to introduce novel coding mechanisms to reduce the time needed to fulfill both UL and DL steps.

The coding solutions devised in this paper draw inspiration from the coded caching (CC) technique, originally proposed to reduce the load at peak traffic times by employing the caches distributed in the network as a supplementary communication resource~\cite{maddah2014fundamental}. In a single-stream downlink network with $K$ users, each with sufficient memory to store a fraction $\gamma$ of the entire file library, CC boosts the achievable rate by the multiplicative factor of $K\gamma + 1$, which scales with the cumulative cache size in the entire network. This {new gain} is accomplished by multicasting carefully designed codewords to different subsets of users with size $K\gamma + 1$, and can also be aggregated with the spatial multiplexing gain to enable the speed-up factor of $K\gamma+L$ in a multi-input multi-output~(MISO) setup with $L$ antennas at the transmitter~\cite{shariatpanahi2016multi,shariatpanahi2018physical}. Due to these exciting properties, CC has been extensively studied in the literature, to address its challenges, such as exponentially growing subpacketization~\cite{salehi2020lowcomplexity,lampiris2018adding}, complex beamformer design~\cite{tolli2017multi,salehi2019subpacketization}, privacy~\cite{9249022}, and applicability to dynamic setups~\cite{10206827}, and to investigate its benefits in use cases such as large-scale video-on-demand (VoD)~~\cite{9361706} and extended reality (XR)~\cite{salehi2022enhancing,mahmoodi2021non}. Variations of the original CC models have also been studied, e.g., for data shuffling~\cite{8957508,8955811,8754795} and linear function retrieval~\cite{9378566,9681716}.

In the context of coded MIR, existing works in the literature have primarily considered the application of CC in distributed computing systems, which offer various advantages, such as enhanced scalability, reliability, and cost-effectiveness, over centralized computing solutions~\cite{book1}. In such systems, the setup mainly consists of a network of $K$ computing nodes, 
a library of files $\{d^n\}_n$, and a set of functions $\{\phi^n(\cdot)\}_n$. Each file $d^n$ is split into non-overlapping portions $d^n_{\CP}$, where $\CP$ can be any subset of computing nodes with a predefined size. Each computing node~$k$ calculates the output of all the functions over all the file parts $\{d^n_{\CP}\}_n$ for which $k \in \CP$. Subsequently, the nodes exchange their calculated outputs, ensuring that each node $k$ could ultimately reconstruct the output of its desired function $\phi^k(d^k)$ over its specified file $d^k$. Coding mechanisms resembling those of CC have been introduced to alleviate the communication load during the data exchange, where the excess computed elements at each node are used to remove undesired terms from the received signals, similar to cache-aided interference removal. The result is a balance between the excess computation power and the required communication load among servers~\cite{7935426,8744394,7954644,9834485}. 

In this paper, 
we propose new coding schemes, inspired by CC, for
wireless MIR scenarios where the information exchange among users is assisted by a multi-antenna BS as the helper node (similar to a two-way relay setup~\cite{4401421}).
With the proposed solution,
the aim is to generate the multicast signals in the DL step based on the signals received during the UL step, such that the DL step performs similarly to the downlink communication of the original MISO-CC scheme~\cite{shariatpanahi2016multi,shariatpanahi2018physical}.
For this purpose, a novel transmission design is required for the UL step, benefiting from the over-the-air addition of the signals transmitted by network users and the modest computation capability at the BS to create the codewords needed in the DL step over a small number of transmission slots. We show that the time required in the UL step can, in fact, be made equal to the time required in DL, which has already been shown to be information-theoretically optimal under simple conditions in downlink MISO-CC communications~\cite{9663389}.
%
As multi-antenna connectivity is an integral part of all modern communication systems, including 5G and beyond cellular networks~\cite{rajatheva2020white}, the coding solutions devised in this paper for MIR are applicable to diverse scenarios, 
such as industrial IoT (exchanging measurements or observations)~\cite{7589556}, distributed coded computing (exchanging computation results)~\cite{7935426}, or distributed cache networks (exchanging cached contents)~\cite{maddah2014fundamental}.


\textit{Notation:} In this paper, bold lower-case and calligraphic letters show vectors and sets, respectively. Moreover, $\Bv[j]$ represents the $j$-entry of vector $\Bv$. We use $\BM^{\rm T}$ and $\BM \herm$ to demonstrate the transpose and conjugate-transpose (Hermitian) of matrix $\BM$, respectively. For integers $a$ and $b$, $\left[ a:b \right]$ shows the set $\lbrace a,\cdots,b \rbrace$ and $[a]=\lbrace 1,\cdots,a \rbrace$.  $\left\vert \CA \right\vert$ is the cardinality of $\CA$, and for $\CB \subseteq \CA$, $\CA\backslash \CB$ represents $\{x\in \CA: x\notin \CB\}$. 

\section{Problem Formulation}

\begin{figure}[t]
    \centering
    \includegraphics[scale=.3]{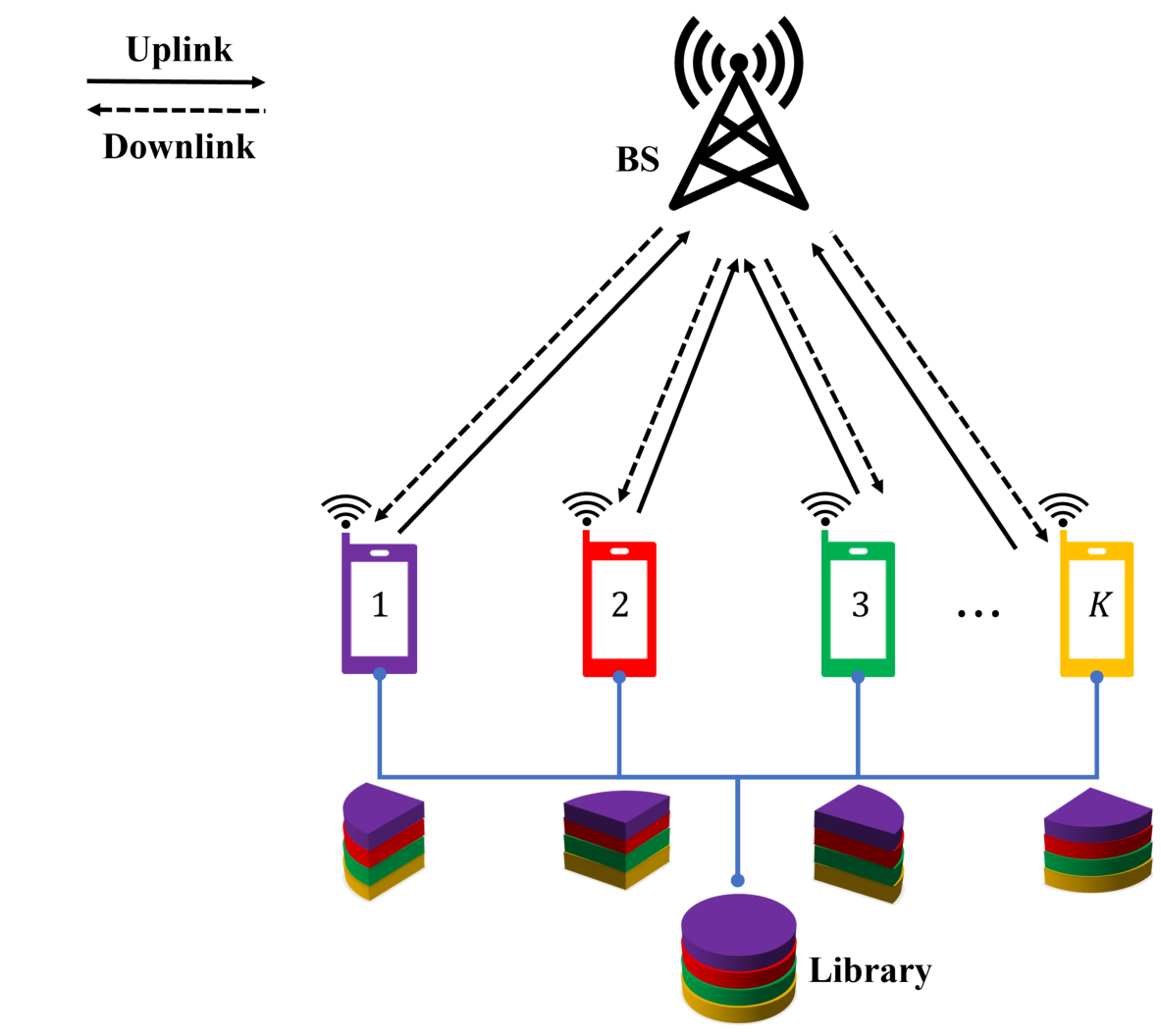}
    \caption{System model for a coded data retrieval network with $K$ single antenna users and a BS with $L$ antennas: A $\gamma$ portion of the entire library is either locally generated or stored beforehand in the cache memories of users. 
    }
    \label{fig: system}
\end{figure}

This paper aims to leverage the underlying coding mechanism of CC to improve multi-user information retrieval in application scenarios where each user has direct access only to a part of each content (i.e., data files), 
but requires all parts of one or more specific contents. We emphasize that in this context, the content may be generated online (e.g., in industrial IoT applications)~\cite{7589556} or be cached in advance~\cite{maddah2014fundamental}. 
In order to exchange data, as depicted in Fig.~\ref{fig: system}, $K$ single-antenna users communicate with an $L$-antenna BS. Here, we consider the worst-case scenario where each user is required to recover a distinct content file. As the contents are only partially accessible by each user and the BS lacks direct access to the content library to satisfy users' demands, a relay-type two-way UL-DL model is used for data exchange. In the UL step, users transmit a portion of their accessible contents to the BS via a number of consecutive UL transmissions. Assuming the BS possesses sufficient computation capability to process and enough memory to store all received signals, it then appropriately processes and combines the received signals, and forwards them back to the users in the DL step.\footnote{In the context of distributed coded computing, the proposed model can be considered as a multi-antenna extension of the system models described in~\cite{8051074,7935426}.}

The content library consists of $N$ files $W^n$, $n \in [N]$, each with a size of $F$ bits. Without loss of generality, we assume $N=K$. 
Let us use $W^n_{\CP}$, $\CP \subseteq [K]$, to denote the part of the content file $W^n$ that can be accessed by every user in $\CP$. We impose two simplifying assumptions: 1) each part of each file is accessible by the same number of users $t$, i.e., $W^n_{\CP}$ has a size larger than zero bits if and only if $|\CP| = t$, and 2) all the file parts $W^n_{\CP}$ with a size larger than zero bits have the same size of $F/\binom{K}{t}$ bits. 
With these assumptions, the system is analogous to a distributed MISO cache network where each cache is sufficiently large to store a fraction $\gamma$ of the entire library, and the CC gain is $t=K\gamma$. As a result, for the sake of simplicity, we may use the standard notation of MISO-CC systems: we use \emph{packet} to denote a part of a content file and say a packet is \emph{cached} by a user if the user has access to the respective file part. Respectively, the above-mentioned assumptions on the access of the users to content files
could also be described by the so-called \emph{content placement phase} of a MISO-CC system, where each content file $W^{n}$, $n \in [N]$, is split into~$\binom{K}{t}$ equal-sized disjoint packets $W^{n}_{\CP}$ as  
\begin{equation}
    W^{n} = \left\lbrace W^{n}_{\CP}: \CP \subseteq \left[ K \right], \left\vert \CP \right\vert=t \right\rbrace,
\end{equation}
and for each $\CP \subseteq [K]$ with $\left\vert \CP \right\vert=t$, each user $k \in \CP$ stores $W^{n}_{\CP}$ in its cache memory. Clearly, the total number of packets stored by each user $k \in [K]$ is $K \binom{K-1}{t-1} = t \binom{K}{t} = \gamma K \binom{K}{t} $, satisfying the cache size constraint.

Without loss of generality (up to a permutation of the indices of the users), let us assume that user $k \in [K]$ requires the content file $W^{k}$. To initiate data exchange, we first further split each packet $W^{n}_{\CP}$ into $\binom{K-t-1}{L-1}$ equal-sized \emph{subpackets} $W^{n}_{\CP, q}$ as 
        $W^n_{\CP} = \lbrace W^n_{\CP,q}: q \in [ \binom{K-t-1}{L-1} ]  \rbrace$,
where $W^n_{\CP,q}$ is comprised of $f$ uniformly i.i.d. bits with $f=\nicefrac{F}{\binom{K}{t} \binom{K-t-1}{L-1}}$.

\begin{defn}
\label{defintion: codeword}
For some $\CV \subseteq [K]$ and an arbitrary $\theta^{k}_{\CP_{k}, q_{k}} \in \mathbb{C}$, the superposition signal 
   \begin{equation}
   \sum_{k \in \CV} \theta^{k}_{\CP_{k}, q_{k}} \Sfc \left( W^{k}_{\CP_{k}, q_{k}} \right),
   \end{equation}
is defined as a \emph{codeword} of size $\left\lvert \CV \right\rvert$, where $\CP_{k} \subseteq [K]$ with $\left\lvert \CP_{k} \right\rvert=t$, $q_{k} \in [\binom{K-t-1}{L-1}]$, and 
$\Sfc \left( W^{k}_{\CP_{k}, q_{k}} \right) $ is the encoded signal of $ W^{k}_{\CP_{k}, q_{k}}$ with $\Sfc: \mathbb{F}_{2^{f}} \rightarrow \mathbb{C}$. 
\end{defn}

Assume that user $k \in [K]$ is connected to the BS through the channel $\Bh_{k} \in \mathbb{C}^{L\times 1}$.
In this paper, we assume that the wireless links are semi-static and channel state information (CSI) is available at the BS and the users.
The UL step comprises $N_{\rm UL}$ transmissions, where in transmission $j \in [N_{\rm UL}]$, each user in a subset $\CU (j)$ of users transmits a fraction of its cache contents to the BS. Hence, after the transmission $j$, the BS receives the signal
\begin{equation}
\label{eq: system UL def}
    \begin{aligned}
        \By_{\mathrm{BS}}(j) = \sum_{k \in \CU ({j})} \Bh_{k} x_{k} ({j}) + \Bn_{\rm BS},
    \end{aligned}
\end{equation}
where $\Bn_{\rm BS} \sim \CC \CN \left(0, \BI_{L} \right)$, and $x_{k}({j})$ is the transmitted signal of user $k$ with power $P_{k}({j})=\mathbb{E} [ \left\vert x_{k}({j}) \right\vert^{2} ]$. Here, it is assumed that the average transmit power of user $k \in [K]$ during the UL step is equal to $P_{\rm T}$, i.e.,  $\frac{1}{N_{\rm UL}} \sum_{j=1}^{N_{\rm UL}} P_{k}({j}) =P_{\rm T} $. 

The DL step involves $N_{\rm DL}$ consecutive transmissions, such that in transmission $j \in [N_{\rm DL}]$, the BS transmits a superposition signal containing the required codewords for a specific set of users represented by $\CV ({j})$.
Therefore, the received signal at user $k \in \CV ({j})$ in the $j$-th DL transmission can be written as
\begin{equation}
\label{eq: y k system def}
    \begin{aligned}
        y_{k}(j)=  \sum\limits_{k \in \CV ({j})} \Bh_{k}^{\rm H}  \Bs_{k}({j}) + n_{k},
    \end{aligned}
\end{equation}
where $\Bs_{k} ({j}) \in \mathbb{C}^{L\times 1}$ is the precoded signal for user $k$ during transmission $j$, $n_{k} \sim \CC \CN \left(0, 1 \right)$, $P_{\rm BS}$ is the BS transmit power in each transmission, such that  $P_{\rm BS} = \sum_{k \in \CV ({j})} \mathbb{E} [ \left\Vert \Bs_{k}({j}) \right\Vert^{2} ]$.

In this paper, the aim is to design a UL-DL transmission strategy, minimizing the \emph{normalized delivery time} (NDT) in the UL and DL steps at high-SNR regimes, i.e., when $P_{\rm T} \rightarrow \infty$ and $P_{\rm BS} \rightarrow \infty$. By following a similar approach presented in~\cite{7954644} and \cite{9834485}, the NDT for the UL step is defined as 
\begin{equation}
\label{eq: TUL highSNR def.}
    \begin{aligned}
        T_{\rm UL} = \lim_{\mathrm{SNR} \rightarrow \infty} \sum\limits_{i=1}^{N_{\rm UL}} \frac{ D_{i}^{\rm UL} d}{\nicefrac{F}{ \log (\mathrm{SNR})}},
    \end{aligned}
\end{equation}
where $ D_{i}^{\rm UL} = \frac{FQ}{R_{i}}$ is the transmission time to deliver data in the $i$-th UL transmission with $R_{i}$ expressing the achievable rate of the user with the worst channel condition, and $Q=\nicefrac{1}{\binom{K}{t} \binom{K-t-1}{L-1}}$. Moreover, $d$ is the per-user degree-of-freedom (DoF) at high SNR~\cite{7954644}. Here, we note that $\nicefrac{F}{ \log (\mathrm{SNR})}$ is the transmission time of delivering a single file of $F$ bits in a single-antenna point-to-point baseline system with Gaussian noise at the high-SNR regime. Considering the high-SNR condition, all optimal transmit and receive beamformers asymptotically behave as zero-forcing (ZF) precoders, and consequently, we have $R_{i} \approx d \log (\mathrm{SNR}) + \CO \left( \log (\mathrm{SNR}) \right) $. As a result, the NDT expression in \eqref{eq: TUL highSNR def.} is simplified to
\begin{equation}
\label{eq: T UL def.}
    \begin{aligned}
        T_{\rm UL} & = \lim_{\mathrm{SNR} \rightarrow \infty}  \frac{ N_{\rm UL} FQ d \log (\mathrm{SNR}) }{ \left( d \log (\mathrm{SNR}) + \CO \left( \log (\mathrm{SNR}) \right) \right)  F}  =  N_{\rm UL} Q.
    \end{aligned}
\end{equation}
Following the same process for the DL step,  
the NDT is expressed as
\begin{equation}
\label{eq: T DL def.}
    \begin{aligned}
        T_{\rm DL} = N_{\rm DL} Q.
    \end{aligned}
\end{equation}

\section{Reference Strategies}
\label{section: naive}
In this section, we introduce two baseline  UL-DL transmission strategies, that leverage either the spatial multiplexing gain $L$ or the CC gain $t$ during the UL step while adopting a transmission approach similar to the MISO-CC scheme of~\cite{shariatpanahi2016multi,shariatpanahi2018physical} for the DL step.

\subsubsection{Strategy A} 
As mentioned, a $\gamma$ portion of the entire library is generated/stored by each user. Therefore, $K (1-\gamma) \binom{K}{t} \binom{K-t-1}{L-1}$ subpackets must be transmitted to the BS in the UL step. To this end, during each UL transmission in \emph{Strategy A}, we select $L$ users to simultaneously transmit  $L$ subpackets to the BS, which is then able to decode all of them as it is equipped with $L$ antennas. Therefore, during the UL step, this strategy only benefits from the spatial multiplexing gain of $L$  without incorporating the CC gain. Hence, employing \eqref{eq: T UL def.}, the NDT in the UL step via \emph{Strategy~A} is given by:
\begin{equation}
\label{eq: R UL naive 1}
    T^{A}_{\rm UL} =  \frac{K \left( 1-\gamma \right)\binom{K}{t} \binom{K-t-1}{L-1}}{L \binom{K}{t} \binom{K-t-1}{L-1}}= \frac{K-t}{L}.
\end{equation}

Following the UL step, the BS has access to all the missing subpackets. As stated in~\cite{9663389}, the optimal transmission strategy among all linear one-shot schemes with uncoded placement for the DL step is to follow a similar approach as the MISO-CC scheme of~\cite{shariatpanahi2016multi,shariatpanahi2018physical}. With this scheme, using \eqref{eq: T DL def.}, the NDT in the DL step of \emph{Strategy A} is $ T^{A}_{\rm DL} = \frac{K-t}{t+L}$.

 \subsubsection{Strategy B}
With this strategy, in each UL transmission,
$t+1$ users are chosen to transmit $t+1$ subpackets to the BS simultaneously. The data transmitted by these users is then added over the air to form one of the codewords needed in the following DL step (more details are provided shortly after). Hence, during the UL step, this strategy only benefits from the CC gain without using the available spatial multiplexing gain. According to~\eqref{eq: T UL def.}, the UL step via \emph{Strategy B} achieves the NDT 
 \begin{equation}
     \begin{aligned}
         T^{B}_{\rm UL} = \frac{K \left( 1-\gamma \right) \binom{K}{t} \binom{K-t-1}{L-1}}{\left( t+1 \right)  \binom{K}{t} \binom{K-t-1}{L-1}} = \frac{K-t}{t+1}.
     \end{aligned}
 \end{equation}

For the DL step, in order to optimize the transmission delay, the BS follows a similar scheme as the one proposed in~\cite{shariatpanahi2016multi,shariatpanahi2018physical}. To this end, for each $\CT \subseteq [K]$ with $\left\vert \CT \right\vert = t+L$, the BS has received $\binom{t+L}{t+1}$ signals during the UL step. Therefore, it generates $\binom{t+L-1}{t}$ random linear combinations of these $ \binom {t+L}{t+1}$ received signals and broadcasts them sequentially. For the decoding process,
we adopt the signal-level decoding approach as expressed in~\cite{lampiris2018adding}, where the undesired terms are regenerated from the local memory and removed before the received signal is decoded by the users. The details of the DL step are presented in Section~\ref{section: DL step}. Accordingly, utilizing \eqref{eq: T DL def.}, the NDT of the DL step via \emph{Strategy B} is obtained as $ T^{B}_{\rm DL} = \frac{K-t}{t+L}$.

As observed, both strategies achieve the same NDT during the DL step; however, in the UL step, neither \emph{Strategy A} nor \emph{Strategy~B} achieves the NDT of $\frac{K-t}{t+L}$. In this work, we devise a UL transmission strategy that incorporates both spatial multiplexing and CC gains to achieve a UL NDT equal to that of the DL.

\section{The New UL-DL Transmission Strategy}
\label{section:new_scheme}
We first present an illustrative example to give further insight into the system performance and then design the generalized transmission strategies for the UL and DL steps. 

\subsection{An Illustrative Example}
Consider a cache-aided MISO network with $K=3$ users, $\gamma=\frac{1}{3}$, $t=1$ and $L=2$. Users $1$, $2$ and $3$ are required to recover the files $W^{1}$, $W^{2}$ and $W^{3}$, respectively. First, each file $W^{n}$ is split into $\binom{K}{t}=3$ packets $W^{n}_{\CP}$, where $\CP \in [3]$. Accordingly, user $k \in [3]$ stores the packets $W^{n}_{k}$ in its cache memory for all $n \in [3]$. 
For simplicity, let us use $A$, $B$, and $C$ to denote the encoded signals of $W^1$, $W^2$, and $W^3$, respectively. Moreover, as $q=\binom{K-t-1}{L-1}=1$, we ignore the index $q$.

The UL step is comprised of $N_{\rm UL}=2$ transmissions, represented by $\lbrace 12 \rbrace$ and $\lbrace 13 \rbrace$. During the transmission $ \CS \in \left\lbrace \lbrace 12 \rbrace, \lbrace 13 \rbrace \right\rbrace$, user $k \in [3]$ transmits the signal $x^{k} (\CS )$, as shown in Table~\ref{table: UL simple exmp}. Assuming noise-less channels,\footnote{The generalized UL and DL steps with noisy-channels are discussed in Section~\ref{section: UL A} and Section~\ref{section: DL step}.} the received signal of the BS during transmission $\CS$, denoted by $\By_{\rm BS} (\CS)$ is given by:
\begin{equation*}
    \begin{aligned}
    \label{eq: y12}
       & \By_{\rm BS} (\lbrace 12 \rbrace ) = \Bh_{1} B_{1} + \Bh_{2} A_{2} + \Bh_{3} \left( -B_{3} - A_{3} \right),\\
       & \By_{\rm BS} (\lbrace 13 \rbrace) = \Bh_{1} C_{1} + \Bh_{2} \left( -C_{2} - A_{2} \right) + \Bh_{3} A_{3}.
    \end{aligned}
\end{equation*}
%
However, in the upcoming DL step, the BS only needs
$\alpha_{1} B_{1} + \alpha_{2} A_{2} $, $\alpha_{3} C_{1} + \alpha_{4} A_{3} $, and $\alpha_{5} C_{2} + \alpha_{6} B_{3} $, for some $\alpha_{i} \in \mathbb{C}$ with $i \in [6]$. 
To extract these combinations, it employs receive beamforming (row) vectors $\Bv_{\CR} \in \mathbb{C}^{1 \times 2}$, such that $\CR \subset [3]$ with $\left\vert \CR \right\vert = t+1 =2$, that satisfy
%
\begin{equation*}
    \begin{aligned}
        \begin{cases}
            \Bv_{\CR}  \Bh_{k} = 0 & k \in [3] \text{ and } k \notin \CR \\
            \Bv_{\CR}  \Bh_{k} \neq 0 & k \in [3] \text{ and } k \in \CR
        \end{cases}.
    \end{aligned}
\end{equation*}
First, using $\Bv_{\lbrace 12 \rbrace}$ and $\Bv_{\lbrace 13 \rbrace}$, the BS can compute
\begin{equation*}
    \begin{aligned}
       & \Bv_{\lbrace 12 \rbrace} \By_{\rm BS} (\lbrace 12 \rbrace) =  \Bv_{\lbrace 12 \rbrace} \Bh_{1} B_{1} +  \Bv_{\lbrace 12 \rbrace} \Bh_{2} A_{2}, \\
       &  \Bv_{\lbrace 13 \rbrace} \By_{\rm BS} \left( \lbrace 13 \rbrace \right) =  \Bv_{\lbrace 13 \rbrace} \Bh_{1} C_{1} +  \Bv_{\lbrace 13 \rbrace} \Bh_{3} A_{3},
    \end{aligned}
\end{equation*}
which yield $\alpha_{1} B_{1} + \alpha_{2} A_{2} $ and $\alpha_{3} C_{1} + \alpha_{4} A_{3} $ by setting $\alpha_{1}=\Bv_{\lbrace 12 \rbrace} \Bh_{1}$, $\alpha_{3} = \Bv_{\lbrace 13 \rbrace} \Bh_{1} $, and so on.
Next, to calculate $\alpha_{5} C_{2} + \alpha_{6} B_{3} $, the BS simply uses $\Bv_{\lbrace 23 \rbrace}$ to get
\begin{equation*}
    \begin{aligned}
        \Bv_{\lbrace 23 \rbrace} \left( \By_{\rm BS} (\lbrace 12 \rbrace) \! + \! \By_{\rm BS} (\lbrace 13 \rbrace) \right) \! \!= \!  
        - \Bv_{\lbrace 23 \rbrace} \Bh_{2} C_{2}  - \Bv_{\lbrace 23 \rbrace} \Bh_{3} B_{3}.
    \end{aligned}
\end{equation*}

\begin{table}[t]
\caption{Transmitted Signals During the UL Step}
 \centering
  \begin{tabular}{|c|c|c|c|}
    \hline
    \multicolumn{1}{|c|}{{\diagbox[innerwidth=1.3cm]{$\CS$}{$x^{k} (\CS)$}}} & $x^{1} (\CS)$ & $x^{2} (\CS)$ &  $x^{3} (\CS)$    \\
    \hline
    $\lbrace 12 \rbrace$ & $B_{1}$ & $A_{2}$ & $-B_{3}-A_{3}$ \\
    \hline
    $\lbrace 13 \rbrace$ & $C_{1}$ & $-C_{2}-A_{2}$ &  $A_{3}$   \\
    \hline
  \end{tabular}
  \label{table: UL simple exmp}
\end{table}

Now, let us review the DL step in more detail. It
consists of the following $N_{\rm DL} =2$ transmissions:
\begin{equation*}
    \begin{aligned}
         \Bx_{\rm BS} (j) \!  & =  \!  \beta_{1,j} (  \alpha_1 B_{1} \! \! + \! \alpha_2 A_{2} ) \Bw_{\{ 12 \} } 
         \! + \! \beta_{2,j} \left( \alpha_3 C_{1} \!\! + \!  \alpha_4 A_{3} \right) \Bw_{  \lbrace 13 \rbrace} \\
        &  + \! \beta_{3,j} \left( \alpha_5 C_{2}  \! + \! \alpha_6 B_{3} \right) \Bw_{\lbrace 23 \rbrace}, 
    \end{aligned}
\end{equation*}
where $\{\beta_{i,j}\}$, $i \in [3]$ and $j \in [2]$, is a set of scalars selected by the BS (more explanation is provided shortly), and for $\CQ \subset [3]$ with $\left\vert \CQ \right\vert=2$, $\Bw_{\CQ} \in \mathbb{C}^{2 \times 1}$ is the beamforming vector that suppresses the interference at user $[3] \backslash \CQ$. 
Let us consider the decoding process at user~1. Applying
ZF precoders, in the transmission $j \in [2]$, it observes
\begin{equation*}
    \begin{aligned}
        y_{1} (j)   & =    \beta_{1,j} (  \alpha_1 B_{1}   + \alpha_2 A_{2} )\Bh_1\herm \Bw_{\{ 12 \} } \\
          & +  \beta_{2,j} \left( \alpha_3 C_{1} \!\! + \!  \alpha_4 A_{3} \right) \Bh_1\herm \Bw_{  \lbrace 13 \rbrace}.
    \end{aligned}
\end{equation*}
User~1 is interested in $A_2$ and $A_3$. As it has $B_1$ and $C_1$ in its cache, it can regenerate and remove their interference terms from $y_1(j)$ to get $\hat{y}_1(j)$. Then, using
 \begin{equation}
 \label{eq: exmp_eq}
     \begin{aligned}
        \begin{bmatrix}
            \alpha_2 \Bh_1\herm \Bw_{\{12\}} A_2  \\  \alpha_4 \Bh_1\herm \Bw_{\{13\}} A_3
        \end{bmatrix}
         = 
         \begin{bmatrix}
             \beta_{1,1} & \beta_{2,1}  \\
             \beta_{1,2} &  \beta_{2,2} \ \\
         \end{bmatrix}^{-1}
         \begin{bmatrix}
            \hat{y}_1(1) \\ \hat{y}_1(2)
        \end{bmatrix},
     \end{aligned}
 \end{equation}
 and following a simple decoding step, it can recover $A_2$ and $A_3$ interference-free. Note that the scalars $\{\beta_{i,j}\}$ should be selected such that the matrix in~\eqref{eq: exmp_eq} and similar matrices for other users are invertible  (this can be done, e.g., using a predefined codebook).
%
As observed, both UL and DL steps are comprised of two transmissions, and  $Q= \nicefrac{1}{\binom{K}{t} \binom{K-t-1}{L-1}} = \frac{1}{3}$. Therefore, by using \eqref{eq: T UL def.}, and $\eqref{eq: T DL def.}$, the NDT for the UL and DL steps is given by:
        $T_{\rm UL} = T_{\rm DL} = \frac{2}{3}$.


\subsection{UL Step}
\label{section: UL A}
This step involves $N_{\rm S}= \binom{K}{t+L}$ stages, with each stage having $N_{\rm T}=\binom{t+L-1}{t}$ transmissions.
During each transmission of stage $i \in \left[ N_{\rm S} \right]$, a set of $t+L$ users denoted by $\CU(i)$, 
encode their generated/stored subpackets and transmit them to the BS at the same time. Here, the key idea of the UL transmission strategy is to enable the BS to create the codewords of size $t+1$, required for the DL step. Without loss of generality (up to the permutation of users' indices), let us focus on the UL transmission strategy during stage $1$, where users $\CU(1)=[t+L]$ simultaneously send their encoded data to the BS. Now, for notational simplicity, for any subset $\CT \subseteq \left[ t+L \right]$ and $l \in \CT$, we define the operator $\left< l \right>_{\CT}$ as follows
\begin{equation}
\label{eq: shift def}
    \begin{aligned}
        \left< l \right>_{\CT}= 
    \end{aligned}
    \begin{cases}
        \min \left\lbrace i \in \CT : i >l \right\rbrace & l < \max \left\lbrace i \in  \CT \right\rbrace \\
        \min \left\lbrace i \in \CT \right\rbrace & l = \max \left\lbrace i \in  \CT \right\rbrace
    \end{cases}.
\end{equation}
As per \eqref{eq: shift def}, $\left< l \right>_{\CT}$ returns the element in $\CT$ that is next to $l$ (according to a circular shift). For example, for
$\CT=\left\lbrace 1,2,3 \right\rbrace$, we have $\left< 1 \right>_{\CT}=2$, $\left< 2 \right>_{\CT}=3$ and $\left< 3 \right>_{\CT}=1$. In order to create the transmitted signals for users $k \in \left[ t+L \right]$, first, we define 
\begin{equation}
\label{eq: set M def}
    \CM=\left\lbrace \CS: \CS \subseteq \left[ t+L \right],  1  \in \CS, \left\vert \CS \right\vert=t+1 \right\rbrace,
\end{equation}
where $\vert \CM \vert=N_{\rm T}$. As mentioned earlier, stage $1$ is also comprised of $N_{\rm T}$ transmissions. Let us represent the transmitted signal of user $k \in \left[ t+L \right]$ during stage $1$ by $x^{k} ( \CS ) $, where $\CS \in \CM$. In this regard, for each $k \in \left[ t+L \right]$ and $ \CS \in \CM$, $x^{k} ( \CS )$ is given by:
\begin{equation}
\label{eq: x KS A def}
    \begin{array}{c}
        x^{k} ( \CS ) =
    \begin{cases}
      \Sfc \big( W^{\left< k \right>_{\CS}}_{\CS \backslash \lbrace \left< k \right>_{\CS} \rbrace ,q} \big) & k \in \CS \\
      - \sum\limits_{j \in \CS}   \Sfc \Big( W^{ \left< k \right>_{\CS \cup \left\lbrace k \right\rbrace  \backslash \lbrace j \rbrace} }_{ \CS \cup \lbrace k \rbrace \backslash \lbrace j, \left< k \right>_{\CS \cup \lbrace k \rbrace \backslash \lbrace j \rbrace } \rbrace , q} \Big) & k \notin \CS
    \end{cases},
    \end{array}
\end{equation}
where $q \in [\binom{K-t-1}{L-1}]$ increases sequentially after each transmission to ensure none of the subpackets is transmitted twice. As mentioned in Definition~\ref{defintion: codeword}, $ \Sfc \left( W^{n}_{ \CP ,q} \right)$ is the encoded signal of $W^{n}_{\CP ,q}$, where $\Sfc: \mathbb{F}_{2^{f}} \rightarrow \mathbb{C}$. In addition, for all $\CQ \subseteq [t+L]$ with $\left\lvert \CQ \right\rvert = t+1$ and $k \in \left[ t+L \right]$, it is assumed that:
\begin{equation}
    \begin{array}{c}
        \mathbb{E} \left[ \left\vert \Sfc \big( W^{\left< k \right>_{\CQ}}_{\CQ \backslash \lbrace \left< k \right>_{\CQ} \rbrace ,q} \big) \right\vert^{2} \right]=
    \label{eq: power allocation A}
    \begin{cases}
    P_{\rm UL} & k=1 \\
        \frac{P_{\rm UL} N_{\rm T}}{\binom{t+L-2}{t-1} + \left( t+1 \right) \binom{t+L-2}{t}  } & k \neq 1
    \end{cases},
    \end{array}
\end{equation}
where $P_{\rm UL}$ represents the transmit power of user $1$ during each transmission of stage $1$, i.e., $P_{1}(j) = P_{\rm UL} $ for all $j \in [N_{\rm T}]$. As per~\eqref{eq: x KS A def}, during stage $i \in [N_{\rm S}]$, $t+L-1$ users transmit a superposition of $t+1$ encoded subpackets in several transmissions. However, in each stage, there is always one user (e.g., user 1 in stage 1) that transmits a single encoded subpacket to the BS. Accordingly, for any stage $i \in [N_{\rm S}]$, the transmit power of the user that transmits a single encoded subpacket to the BS in all transmissions is fixed at $P_{\rm UL}$.  
In~\eqref{eq: power allocation A}, the expectation operates on the uniformly i.i.d. random variable $\Sfc \big( W^{\left< k \right>_{\CQ}}_{\CQ \backslash \lbrace \left< k \right>_{\CQ} \rbrace ,q} \big)$, as it is assumed that $W^{\left< k \right>_{\CQ}}_{\CQ \backslash \lbrace \left< k \right>_{\CQ} \rbrace ,q}$ is an i.i.d. random variable uniformly distributed on $\mathbb{F}_{2^{f}}$.
\begin{lem}
\label{lemma: power allocation A}
Assuming $E_{k}$ as the energy consumption of user $k \in [K]$ in the UL step, the proposed encoding scheme in~\eqref{eq: power allocation A} satisfies the users' energy constraints $E_{k}=P_{\rm UL} N_{\rm T} \binom{K-1}{t+L-1}$. Hence, although the transmit power of users may vary during each UL transmission, each user consumes an equal total amount of energy throughout all transmissions of the UL step.

\end{lem}
\begin{IEEEproof}
    The proof is relegated to Appendix~\ref{proof: lemma  power allocation}.
\end{IEEEproof}


\begin{defn}
For the subset $\CQ \subseteq [t+L]$ with $\lvert \CQ \rvert = t+1$, define the beamforming (row) vector $\Bv_{\CQ} \in \mathbb{C}^{1 \times L}$ as follows
\begin{equation}
    \begin{aligned}
        \begin{cases}
            \Bv_{\CQ}  \Bh_{k} = 0 & k \in [t+L] \text{ and } k \notin \CQ \\
            \Bv_{\CQ}  \Bh_{k} \neq 0 & k \in \CQ
        \end{cases}.
    \end{aligned}
\end{equation}
\end{defn}

\begin{thm}
\label{theorem: load UL A}
   Using the proposed UL transmission strategy, during stage 1, the BS is able to create all codewords of size $t+1$ in the set $\CA \, \cup \, \CB$, where
   \begin{equation}
    \label{eq: set A def}
        \begin{array}{c}
             \CA  =  \Big\lbrace  \sum\limits_{k \in \CS} \Bv_{\CS} \Bh_{k} \Sfc \big( W^{\left< k \right>_{\CS}}_{\CS \backslash \lbrace  \left< k \right>_{\CS} \rbrace ,q} \big) +  \Bv_{\CS} \Bn_{\rm BS} : \CS \in \CM  \Big\rbrace,
        \end{array}
    \end{equation}
    \begin{equation}
     \label{eq: set B ref}
         \begin{array}{c}
            \CB= \Big\lbrace   \sum\limits_{k \in \CR }  - \Bv_{\CR} \Bh_{k}  \Sfc \big( W^{\left< k \right>_{\CR}}_{\CR \backslash \lbrace  \left< k \right>_{\CR } \rbrace  ,q} \big)  + \Bv_{\CR } \Bn_{\rm BS} :  \\
              \CR \subseteq \left[ 2: t+L \right], \left\vert \CR \right\vert=t+1  \Big\rbrace
         \end{array}
     \end{equation}
    include noisy versions of the codewords of size $t+1$ (that are needed in the DL step).
   Moreover, the NDT for UL is 
   \begin{equation}
   \label{eq: R UL A}
       \begin{aligned}
           T_{\rm UL} = \frac{K-t}{t+L}.
       \end{aligned}
   \end{equation}
\end{thm}

\begin{IEEEproof}
The proof is provided in Appendix~\ref{proof: theorem  UL}.
\end{IEEEproof}


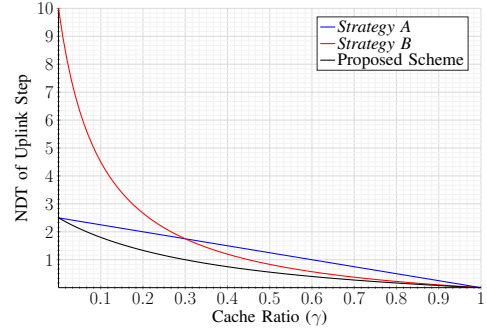
\begin{figure}[t]
   \centering
      \pgfplotstableread[row sep=\\,col sep=&]{
sigmax & MC8\\
0	&	2.5	\\
0.01	&	2.475	\\
0.02	&	2.45	\\
0.03	&	2.425	\\
0.04	&	2.4	\\
0.05	&	2.375	\\
0.06	&	2.35	\\
0.07	&	2.325	\\
0.08	&	2.3	\\
0.09	&	2.275	\\
0.1	&	2.25	\\
0.11	&	2.225	\\
0.12	&	2.2	\\
0.13	&	2.175	\\
0.14	&	2.15	\\
0.15	&	2.125	\\
0.16	&	2.1	\\
0.17	&	2.075	\\
0.18	&	2.05	\\
0.19	&	2.025	\\
0.2	&	2	\\
0.21	&	1.975	\\
0.22	&	1.95	\\
0.23	&	1.925	\\
0.24	&	1.9	\\
0.25	&	1.875	\\
0.26	&	1.85	\\
0.27	&	1.825	\\
0.28	&	1.8	\\
0.29	&	1.775	\\
0.3	&	1.75	\\
0.31	&	1.725	\\
0.32	&	1.7	\\
0.33	&	1.675	\\
0.34	&	1.65	\\
0.35	&	1.625	\\
0.36	&	1.6	\\
0.37	&	1.575	\\
0.38	&	1.55	\\
0.39	&	1.525	\\
0.4	&	1.5	\\
0.41	&	1.475	\\
0.42	&	1.45	\\
0.43	&	1.425	\\
0.44	&	1.4	\\
0.45	&	1.375	\\
0.46	&	1.35	\\
0.47	&	1.325	\\
0.48	&	1.3	\\
0.49	&	1.275	\\
0.5	&	1.25	\\
0.51	&	1.225	\\
0.52	&	1.2	\\
0.53	&	1.175	\\
0.54	&	1.15	\\
0.55	&	1.125	\\
0.56	&	1.1	\\
0.57	&	1.075	\\
0.58	&	1.05	\\
0.59	&	1.025	\\
0.6	&	1	\\
0.61	&	0.975	\\
0.62	&	0.95	\\
0.63	&	0.925	\\
0.64	&	0.9	\\
0.65	&	0.875	\\
0.66	&	0.85	\\
0.67	&	0.825	\\
0.68	&	0.8	\\
0.69	&	0.775	\\
0.7	&	0.75	\\
0.71	&	0.725	\\
0.72	&	0.7	\\
0.73	&	0.675	\\
0.74	&	0.65	\\
0.75	&	0.625	\\
0.76	&	0.6	\\
0.77	&	0.575	\\
0.78	&	0.55	\\
0.79	&	0.525	\\
0.8	&	0.5	\\
0.81	&	0.475	\\
0.82	&	0.45	\\
0.83	&	0.425	\\
0.84	&	0.4	\\
0.85	&	0.375	\\
0.86	&	0.35	\\
0.87	&	0.325	\\
0.88	&	0.3	\\
0.89	&	0.275	\\
0.9	&	0.25	\\
0.91	&	0.225	\\
0.92	&	0.2	\\
0.93	&	0.175	\\
0.94	&	0.15	\\
0.95	&	0.125	\\
0.96	&	0.1	\\
0.97	&	0.075	\\
0.98	&	0.05	\\
0.99	&	0.025	\\
1	&	0	\\
}\AM

\pgfplotstableread[row sep=\\,col sep=&]{
sigmax & MC8\\
0	&	10	\\
0.01	&	9	\\
0.02	&	8.166666667	\\
0.03	&	7.461538462	\\
0.04	&	6.857142857	\\
0.05	&	6.333333333	\\
0.06	&	5.875	\\
0.07	&	5.470588235	\\
0.08	&	5.111111111	\\
0.09	&	4.789473684	\\
0.1	&	4.5	\\
0.11	&	4.238095238	\\
0.12	&	4	\\
0.13	&	3.782608696	\\
0.14	&	3.583333333	\\
0.15	&	3.4	\\
0.16	&	3.230769231	\\
0.17	&	3.074074074	\\
0.18	&	2.928571429	\\
0.19	&	2.793103448	\\
0.2	&	2.666666667	\\
0.21	&	2.548387097	\\
0.22	&	2.4375	\\
0.23	&	2.333333333	\\
0.24	&	2.235294118	\\
0.25	&	2.142857143	\\
0.26	&	2.055555556	\\
0.27	&	1.972972973	\\
0.28	&	1.894736842	\\
0.29	&	1.820512821	\\
0.3	&	1.75	\\
0.31	&	1.682926829	\\
0.32	&	1.619047619	\\
0.33	&	1.558139535	\\
0.34	&	1.5	\\
0.35	&	1.444444444	\\
0.36	&	1.391304348	\\
0.37	&	1.340425532	\\
0.38	&	1.291666667	\\
0.39	&	1.244897959	\\
0.4	&	1.2	\\
0.41	&	1.156862745	\\
0.42	&	1.115384615	\\
0.43	&	1.075471698	\\
0.44	&	1.037037037	\\
0.45	&	1	\\
0.46	&	0.964285714	\\
0.47	&	0.929824561	\\
0.48	&	0.896551724	\\
0.49	&	0.86440678	\\
0.5	&	0.833333333	\\
0.51	&	0.803278689	\\
0.52	&	0.774193548	\\
0.53	&	0.746031746	\\
0.54	&	0.71875	\\
0.55	&	0.692307692	\\
0.56	&	0.666666667	\\
0.57	&	0.641791045	\\
0.58	&	0.617647059	\\
0.59	&	0.594202899	\\
0.6	&	0.571428571	\\
0.61	&	0.549295775	\\
0.62	&	0.527777778	\\
0.63	&	0.506849315	\\
0.64	&	0.486486486	\\
0.65	&	0.466666667	\\
0.66	&	0.447368421	\\
0.67	&	0.428571429	\\
0.68	&	0.41025641	\\
0.69	&	0.392405063	\\
0.7	&	0.375	\\
0.71	&	0.358024691	\\
0.72	&	0.341463415	\\
0.73	&	0.325301205	\\
0.74	&	0.30952381	\\
0.75	&	0.294117647	\\
0.76	&	0.279069767	\\
0.77	&	0.264367816	\\
0.78	&	0.25	\\
0.79	&	0.235955056	\\
0.8	&	0.222222222	\\
0.81	&	0.208791209	\\
0.82	&	0.195652174	\\
0.83	&	0.182795699	\\
0.84	&	0.170212766	\\
0.85	&	0.157894737	\\
0.86	&	0.145833333	\\
0.87	&	0.134020619	\\
0.88	&	0.12244898	\\
0.89	&	0.111111111	\\
0.9	&	0.1	\\
0.91	&	0.089108911	\\
0.92	&	0.078431373	\\
0.93	&	0.067961165	\\
0.94	&	0.057692308	\\
0.95	&	0.047619048	\\
0.96	&	0.037735849	\\
0.97	&	0.028037383	\\
0.98	&	0.018518519	\\
0.99	&	0.009174312	\\
1	&	0	\\
}\AT

\pgfplotstableread[row sep=\\,col sep=&]{
sigmax & MC8\\
0	&	2.5	\\
0.01	&	2.414634146	\\
0.02	&	2.333333333	\\
0.03	&	2.255813953	\\
0.04	&	2.181818182	\\
0.05	&	2.111111111	\\
0.06	&	2.043478261	\\
0.07	&	1.978723404	\\
0.08	&	1.916666667	\\
0.09	&	1.857142857	\\
0.1	&	1.8	\\
0.11	&	1.745098039	\\
0.12	&	1.692307692	\\
0.13	&	1.641509434	\\
0.14	&	1.592592593	\\
0.15	&	1.545454545	\\
0.16	&	1.5	\\
0.17	&	1.456140351	\\
0.18	&	1.413793103	\\
0.19	&	1.372881356	\\
0.2	&	1.333333333	\\
0.21	&	1.295081967	\\
0.22	&	1.258064516	\\
0.23	&	1.222222222	\\
0.24	&	1.1875	\\
0.25	&	1.153846154	\\
0.26	&	1.121212121	\\
0.27	&	1.089552239	\\
0.28	&	1.058823529	\\
0.29	&	1.028985507	\\
0.3	&	1	\\
0.31	&	0.971830986	\\
0.32	&	0.944444444	\\
0.33	&	0.917808219	\\
0.34	&	0.891891892	\\
0.35	&	0.866666667	\\
0.36	&	0.842105263	\\
0.37	&	0.818181818	\\
0.38	&	0.794871795	\\
0.39	&	0.772151899	\\
0.4	&	0.75	\\
0.41	&	0.728395062	\\
0.42	&	0.707317073	\\
0.43	&	0.686746988	\\
0.44	&	0.666666667	\\
0.45	&	0.647058824	\\
0.46	&	0.627906977	\\
0.47	&	0.609195402	\\
0.48	&	0.590909091	\\
0.49	&	0.573033708	\\
0.5	&	0.555555556	\\
0.51	&	0.538461538	\\
0.52	&	0.52173913	\\
0.53	&	0.505376344	\\
0.54	&	0.489361702	\\
0.55	&	0.473684211	\\
0.56	&	0.458333333	\\
0.57	&	0.443298969	\\
0.58	&	0.428571429	\\
0.59	&	0.414141414	\\
0.6	&	0.4	\\
0.61	&	0.386138614	\\
0.62	&	0.37254902	\\
0.63	&	0.359223301	\\
0.64	&	0.346153846	\\
0.65	&	0.333333333	\\
0.66	&	0.320754717	\\
0.67	&	0.308411215	\\
0.68	&	0.296296296	\\
0.69	&	0.28440367	\\
0.7	&	0.272727273	\\
0.71	&	0.261261261	\\
0.72	&	0.25	\\
0.73	&	0.238938053	\\
0.74	&	0.228070175	\\
0.75	&	0.217391304	\\
0.76	&	0.206896552	\\
0.77	&	0.196581197	\\
0.78	&	0.186440678	\\
0.79	&	0.176470588	\\
0.8	&	0.166666667	\\
0.81	&	0.157024793	\\
0.82	&	0.147540984	\\
0.83	&	0.138211382	\\
0.84	&	0.129032258	\\
0.85	&	0.12	\\
0.86	&	0.111111111	\\
0.87	&	0.102362205	\\
0.88	&	0.09375	\\
0.89	&	0.085271318	\\
0.9	&	0.076923077	\\
0.91	&	0.06870229	\\
0.92	&	0.060606061	\\
0.93	&	0.052631579	\\
0.94	&	0.044776119	\\
0.95	&	0.037037037	\\
0.96	&	0.029411765	\\
0.97	&	0.02189781	\\
0.98	&	0.014492754	\\
0.99	&	0.007194245	\\
1	&	0	\\
}\AP

\begin{tikzpicture}[scale=.33]

\begin{axis}[%
width=2.1\columnwidth,
height=1.45\columnwidth,
axis lines = center,
xlabel near ticks,
xlabel={\huge Cache Ratio ($\gamma$)},
ylabel={\huge NDT of Uplink Step},
ylabel near ticks,
    grid=both,
    major grid style={line width=.2pt,draw=gray!30},
    grid style={line width=.1pt, draw=gray!10},
    minor tick num=5,
    legend pos = north east,
legend style={ legend cell align=left, align=left, draw=white!15!black},
ticklabel style={font=\huge},
]
\addplot[solid,draw=blue,line width=1.2pt] table[x=sigmax,y=MC8]{\AM};
\addlegendentry{\huge \emph{Strategy A}}

\addplot[solid,draw=red,line width=1.2pt] table[x=sigmax,y=MC8]{\AT};
\addlegendentry{\huge \emph{Strategy B}}

\addplot[solid,draw=black,line width=1.2pt] table[x=sigmax,y=MC8]{\AP};
\addlegendentry{\huge Proposed Scheme}

\end{axis}
\end{tikzpicture}%
      \vspace{-.8em}
    \caption{Comparison between the achievable UL normalized delivery time of the proposed scheme with strategies \emph{A} and \emph{B} in a MISO network with $K=10$ and $L=4$.}
    \label{fig: load}
    \vspace{-1.2em}
\end{figure}

Using Theorem~\ref{theorem: load UL A} and changing the users' indices in the sets $A$ and $B$, for any stage $i \in [N_{\rm S}]$, one can simply show that the BS is able to create all codewords of size $t+1$, required for the DL step. Fig.~\ref{fig: load} compares the achievable UL NDT of the proposed scheme with the reference strategies in Section~\ref{section: naive}.

\subsection{DL Step}
\label{section: DL step}
The DL step, inspired by \cite{shariatpanahi2016multi,shariatpanahi2018physical}, is comprised of $N_{\rm S}$ stages, each with $N_{\rm T}$ DL transmissions.
Similar to Section~\ref{section: UL A}, we focus on stage $1$, where the BS serves users $\CU (1)=[t+L]$.
During stage 1, for each $\CQ \subseteq [t+L]$ with $\left\vert \CQ \right\vert=t+1$, the BS generates a random vector $\Ba_{\CQ} \in \mathbb{R}^{1 \times N_{\rm T}}$. Then, using \eqref{eq: set A def} and \eqref{eq: set B ref}, in transmission $j \in \left[ N_{\rm T} \right]$, it broadcasts  a superposition of the codewords in the sets $\CA$ and $\CB$ as follows 
\begin{equation}
\label{eq: XBS j A}
    \begin{aligned}
        & \Bx_{\mathrm{BS}} (j) = \sum\limits_{\substack{\CQ  \subseteq [t+L],  \left\vert \CQ \right\vert=t+1}}  \Ba_{\CQ}[j]  f_{\CQ} \Bw_{\CQ},
    \end{aligned}
\end{equation}
where $f_{\CQ}=  \sum\nolimits_{l \in \CQ} \theta_{\CQ} \Bv_{\CQ} \Bh_{l} \Sfc \big( W^{\left< l \right>_{\CQ}}_{\CQ \backslash \lbrace  \left< l \right>_{\CQ} \rbrace ,q} \big) +  \Bv_{\CQ} \Bn_{\rm BS}$, such that $\theta_{\CQ} =1$ if $1 \in \CQ$ and $\theta_{\CQ} = -1$ otherwise. 
Moreover, $\Bw_{\CQ} \in \mathbb{C}^{L\times 1}$ is the precoder that suppresses the interference at the set $[t+L]\backslash \CQ$.

\begin{thm}
\label{theorem: DL}
    By adopting the introduced DL transmission strategy, each user is able to retrieve its requested file, and the achievable NDT for the DL step takes the form of:
    \begin{equation}
   \label{eq: R DL A}
       \begin{aligned}
           T_{\rm DL} = \frac{K-t}{t+L}.
       \end{aligned}
   \end{equation}
\end{thm}
\begin{IEEEproof}
The proof is relegated to Appendix~\ref{proof: DL}.
\end{IEEEproof}

In Appendix~\ref{exmp: general}, we present an example to give further insight into the system performance for the proposed scheme.
\vspace{-1.4em}
\section{Conclusion}
This paper introduced a novel communication strategy to enhance distributed information retrieval in a network comprising multiple users. To design an inclusive setup supporting various user access scenarios, we modified the operational framework of the conventional coded caching (CC) model. In this modified CC model, all types of user access were considered as data partially stored in the cached content of users, while the multi-antenna base station (BS) is not directly connected to the library. The missing data portions were exchanged among users through an uplink (UL) step followed by a downlink (DL) step. During the UL step, users transmitted a smart combination of their generated/cached contents to the BS. In the DL step, an appropriate combination of these signals was forwarded back to the users, enabling each user to retrieve its desired data. It was shown that the UL step achieved an UL delivery time equal to that of the DL step. 
For future work, we aim to prove the optimality of the proposed scheme.



\clearpage



\clearpage

\appendices
\section{Proof of Lemma~\ref{lemma: power allocation A}}
\label{proof: lemma  power allocation}
    In order to prove Lemma~\ref{lemma: power allocation A}, we focus on stage 1, where users $\CU(1)=\left[ t+L \right]$ transmit their encoded cached data. As per~\eqref{eq: x KS A def} and \eqref{eq: power allocation A}, for each $\CQ \subseteq [t+L]$ with $\left\vert \CQ \right\rvert =t+1$ and $1 \in \CQ$, user $1$ transmits $\Sfc \left( W^{\left< 1 \right>_{\CQ}}_{\CQ \backslash \left\lbrace \left< 1 \right>_{\CQ} \right\rbrace ,q} \right)$ with the transmit power $\mathbb{E} \left[ \left\vert \Sfc \left( W^{\left< 1 \right>_{\CQ}}_{\CQ \backslash \left\lbrace \left< 1 \right>_{\CQ} \right\rbrace ,q} \right) \right\vert ^{2} \right]=P_{\rm UL}$. 
     Therefore, the energy consumption of user $1$ during $N_{\rm T}$ transmissions of stage 1 is obtained as $E_{1}=P_{\rm UL}N_{\rm T}$. 
    
    According to \eqref{eq: x KS A def} and \eqref{eq: power allocation A}, for $k \in \left[2:t+L \right]$, and $\CQ \subseteq [t+L]$ satisfying $\left\vert \CQ \right\rvert =t+1$ and $1, k \in \CQ$, user $k $ transmits $\Sfc \left( W^{\left< k \right>_{\CQ}}_{\CQ \backslash \left\lbrace \left< k \right>_{\CQ} \right\rbrace ,q} \right)$ with the transmit power $ \frac{P_{\rm UL} N_{\rm T}}{\binom{t+L-2}{t-1} + \left( t+1 \right) \binom{t+L-2}{t}  }$. Moreover,  for $k \in \left[2:t+L \right]$, and $\CQ \subseteq [t+L]$ satisfying $\left\vert \CQ \right\rvert =t+1$, $1 \in \CQ$, and $k \notin \CQ$, user $k$ transmits the signal $- \sum\nolimits_{j \in \CQ} \Sfc \left( W^{ \left< k \right>_{\CQ \cup \left\lbrace k \right\rbrace  \backslash \left\lbrace j \right\rbrace} }_{ \CQ \cup \left\lbrace k \right\rbrace \backslash \left\lbrace j, \left< k \right>_{\CQ \cup \left\lbrace k \right\rbrace \backslash \left\lbrace j \right\rbrace } \right\rbrace , q} \right)$ with the transmit power $\frac{ \left( t+1 \right) P_{\rm UL}N_{\rm T}}{\binom{t+L-2}{t-1} + \left( t+1 \right) \binom{t+L-2}{t}  }$. As a result, the energy consumption of user $k \in [2: t+L]$ through $N_{\rm T}$ transmissions of stage 1 is obtained as:
    \begin{equation}
        \begin{aligned}
            {E}_{k} &=   \frac{P_{\rm UL}N_{\rm T}}{\binom{t+L-2}{t-1}+ \left( t+1 \right) \binom{t+L-2}{t} }  \\
            & \times  \left[ \binom{t+L-2}{t-1} + \left( t+1 \right) \binom{t+L-2}{t}  \right]=P_{\rm UL} N_{\rm T}.
        \end{aligned}
    \end{equation}
    Accordingly, since each user $k \in [K]$ transmits data to the BS in $\binom{K-1}{t+L-1}$ stages, the total energy consumption of user $k$ in the UL step is given by $E_{k} = P_{\rm UL} N_{\rm T} \binom{K-1}{t+L-1}$, satisfying the users' energy constraints.

\begin{figure*}[htb]
     \begin{equation}
     \label{eq: set B def}
         \begin{aligned}
            \CB= \left\lbrace  \sum\limits_{k \in \CS \cup \left\lbrace l \right\rbrace  \backslash \left\lbrace 1 \right\rbrace  }  - \Bv_{\CS \cup \left\lbrace l \right\rbrace \backslash \left\lbrace 1 \right\rbrace} \Bh_{k}  \Sfc \left( W^{\left< k \right>_{\CS \cup \left\lbrace l \right\rbrace \backslash \left\lbrace 1 \right\rbrace}}_{\CS \cup \left\lbrace l \right\rbrace \backslash \left\lbrace 1,  \left< k \right>_{\CS \cup \left\lbrace l \right\rbrace \backslash \left\lbrace 1 \right\rbrace } \right\rbrace  ,q} \right)  + \Bv_{\CS \cup \left\lbrace l \right\rbrace \backslash \left\lbrace 1 \right\rbrace } \Bn_{\rm BS} : \CS \in \CM, l \in \left[ t+L \right], l \notin \CS \right\rbrace
         \end{aligned}
     \end{equation}
     \hrulefill
     \end{figure*}

\section{Proof of Theorem~\ref{theorem: load UL A}}
\label{proof: theorem  UL}
Here, first, we show that for stage 1, the BS can generate all codewords of size $t+1$ in the set $\CA \, \cup \, \CB$, and then, for the achievable NDT, we prove \eqref{eq: R UL A}. As mentioned in Section~\ref{section: UL A}, during stage 1 of the UL step, users $\CU(1)=\left[ t+L \right]$ transmit their encoded data to the BS. 
In order to create the codewords of size $t+1$ in the set $\CA$, for any $\CS \in \CM$, the BS considers the received signal $\By_{\rm BS}(\CS)$, which by using~\eqref{eq: system UL def}, it is expressed as follows
\begin{equation}
\label{eq: yBS A}
   \begin{aligned}
       \By_{\rm BS} (\CS)=\sum\limits_{k \in \left[ t + L \right]} \Bh_{k} x^{k} \left( \CS \right)  +\Bn_{\rm BS}.
   \end{aligned}
\end{equation}
Then, the BS uses the beamforming (row) vector $\Bv_{\CS} \in \mathbb{C}^{1 \times L}$, such that $\Bv_{\CS} \Bh_{j} \neq 0$ for $j \in \CS$, and $\Bv_{\CS} \Bh_{j}=0$ for $j \notin \CS$ and $j \in [t+L]$.
Accordingly, for any $\CS \in \CM$, by substituting \eqref{eq: x KS A def} into \eqref{eq: yBS A}, it is found that:
\begin{equation}
\label{eq: codeword UL set A}
    \begin{aligned}
        \Bv_{\CS} \By_{\rm BS} (\CS) &  = \! \sum\limits_{k \in \left[ t + L \right]}  \Bv_{\CS}  \Bh_{k} x^{k} (\CS ) +\Bv_{\CS}\Bn_{\rm BS} \\
        & = \sum\limits_{k \in \CS} \Bv_{\CS} \Bh_{k} \Sfc \left( W^{\left< k \right>_{\CS}}_{\CS \backslash \left\lbrace  \left< k \right>_{\CS} \right\rbrace ,q} \right) +  \Bv_{\CS} \Bn_{\rm BS},
    \end{aligned}
\end{equation}
which is a codeword available in the set $\CA$. 

In order to create the codewords in the set $\CB$, 
by using~\eqref{eq: set M def}, \eqref{eq: set B ref} is rewritten as \eqref{eq: set B def} shown in the top of this page. Then, for $l \in [t+L]$, $\CS \in \CM$, and $l \notin \CS$, the BS first uses the beamforming vector $\Bv_{\CS \cup \lbrace l \rbrace \backslash \lbrace 1 \rbrace } \in \mathbb{C}^{1 \times L} $, satisfying
\begin{equation}
\label{eq: h perp Sl}
    \begin{cases}
        \Bv_{\CS \cup \lbrace l \rbrace \backslash \lbrace 1 \rbrace } \Bh_{j} \neq 0 & j  \in \CS \cup \lbrace l \rbrace \backslash \lbrace 1 \rbrace \\
        \Bv_{\CS \cup \lbrace l \rbrace \backslash \lbrace 1 \rbrace } \Bh_{j} = 0 &   j \in [t+L] \text{ and } j  \notin \CS \cup \lbrace l \rbrace \backslash \lbrace 1 \rbrace
    \end{cases}.
\end{equation}
Then, for $l \in [t+L]$, $\CS \in \CM$, and $l \notin \CS$, the BS uses the set of received signals $\left\lbrace \By_{\rm BS} (\CS \cup \lbrace l \rbrace \backslash \lbrace j \rbrace ) : j \in \CS \cup \lbrace l \rbrace \backslash \lbrace 1 \rbrace \right\rbrace$, and by utilizing \eqref{eq: yBS A} computes
   
\begin{equation}
\label{eq: Ysl code def}
    \begin{aligned}
        &\sum\limits_{j \in \CS \cup \lbrace l \rbrace \backslash \lbrace 1 \rbrace} \Bv_{\CS \cup \lbrace l \rbrace \backslash \lbrace 1 \rbrace} \By_{\rm BS} \left( \CS \cup \lbrace l \rbrace \backslash \lbrace j \rbrace \right)=\\
&         \sum\limits_{j \in \CS \cup \lbrace l \rbrace \backslash \lbrace 1 \rbrace} \Bv_{\CS \cup \lbrace l \rbrace \backslash \lbrace 1 \rbrace} \sum\limits_{k \in [t+L]} \Bh_{k} x^{k} \left( \CS \cup \lbrace l \rbrace \backslash \lbrace j \rbrace \right) + \Tilde{n}_{\CS}, 
    \end{aligned}
\end{equation}
where $\Tilde{n}_{\CS}= \Bv_{\CS \cup \lbrace l \rbrace \backslash \lbrace 1 \rbrace} \Bn_{\rm BS}$. Hence, by applying \eqref{eq: h perp Sl}, \eqref{eq: Ysl code def} is simplified to:
\begin{equation}
\label{eq: Ysl code simp}
    \begin{aligned}
&         \sum\limits_{j \in \CS \cup \lbrace l \rbrace \backslash \lbrace 1 \rbrace}  \sum\limits_{k \in \CS \cup \lbrace l \rbrace \backslash \lbrace 1 \rbrace}  \Bv_{\CS \cup \lbrace l \rbrace \backslash \lbrace 1 \rbrace} \Bh_{k} x^{k} \left( \CS \cup \lbrace l \rbrace \backslash \lbrace j \rbrace \right) + \Tilde{n}_{\CS}\\
 & \! \! = \! \! \sum\limits_{k \in \CS \cup \lbrace l \rbrace \backslash \lbrace 1 \rbrace}  \Bv_{\CS \cup \lbrace l \rbrace \backslash \lbrace 1 \rbrace} \Bh_{k} \sum\limits_{j \in \CS \cup \lbrace l \rbrace \backslash \lbrace 1 \rbrace}  \! \! \! x^{k} \left( \CS \cup \lbrace l \rbrace \backslash \lbrace j \rbrace \right) + \Tilde{n}_{\CS}\\
 &  \! \! = \! \! \sum\limits_{k \in \CS \cup \lbrace l \rbrace \backslash \lbrace 1 \rbrace}  \Bv_{\CS \cup \lbrace l \rbrace \backslash \lbrace 1 \rbrace} \Bh_{k}  \Delta_{k} + \Tilde{n}_{\CS},
    \end{aligned}
\end{equation}
where
\begin{equation}
\label{eq: Delta K def}
    \begin{aligned}
    \Delta_{k}= x^{k} \left( \CS \cup \lbrace l \rbrace \backslash \lbrace k \rbrace \right) + \sum\limits_{j \in \CS \cup \lbrace l \rbrace \backslash \lbrace 1,k \rbrace}   x^{k} \left( \CS \cup \lbrace l \rbrace \backslash \lbrace j \rbrace \right).
    \end{aligned}
\end{equation}
Now, by applying \eqref{eq: x KS A def} to \eqref{eq: Delta K def}, and according to that $k \in \CS \cup \lbrace l \rbrace \backslash \lbrace j \rbrace$ for $j \neq k$, $\Delta_{k}$ is obtained as:
\begin{equation}
\label{eq: Deltak}
    \begin{aligned}
 \Delta_{k}= - & \sum\limits_{j \in \CS \cup \lbrace l \rbrace \backslash \lbrace k \rbrace}   \Sfc \left( W^{ \left< k \right>_{\CS \cup \left\lbrace l \right\rbrace  \backslash \left\lbrace j \right\rbrace} }_{ \CS \cup \left\lbrace l \right\rbrace \backslash \left\lbrace j, \left< k \right>_{\CS \cup \left\lbrace l \right\rbrace \backslash \left\lbrace j \right\rbrace } \right\rbrace , q} \right) + \\
 &  \sum\limits_{j \in \CS \cup \lbrace l \rbrace \backslash \lbrace 1,k \rbrace}   \Sfc \left( W^{\left< k \right>_{\CS \cup \lbrace l \rbrace \backslash \lbrace j \rbrace}}_{\CS \cup \lbrace l \rbrace \backslash \left\lbrace j, \left< k \right>_{\CS \cup \lbrace l \rbrace \backslash \lbrace j \rbrace} \right\rbrace ,q} \right) \\
 & \hspace{-2.1em} = - \Sfc \left( W^{\left< k \right>_{\CS \cup \lbrace l \rbrace \backslash \lbrace 1 \rbrace}}_{\CS \cup \lbrace l \rbrace \backslash \left\lbrace 1, \left< k \right>_{\CS \cup \lbrace l \rbrace \backslash \lbrace 1 \rbrace} \right\rbrace ,q} \right).
    \end{aligned}
\end{equation}
Finally, by substituting \eqref{eq: Deltak} and \eqref{eq: Ysl code simp} into \eqref{eq: Ysl code def}, and setting $\CR= \CS \cup \lbrace l \rbrace \backslash \lbrace 1 \rbrace$, it is observed that:
     \begin{equation}
\label{eq: Ysl code final}
    \begin{aligned}
        &\sum\limits_{j \in \CS \cup \lbrace l \rbrace \backslash \lbrace 1 \rbrace}  \Bv_{\CS \cup \lbrace l \rbrace \backslash \lbrace 1 \rbrace} \By_{\rm BS} \left( \CS \cup \lbrace l \rbrace \backslash \lbrace j \rbrace \right) =  \\
        &- \sum\limits_{k \in \CR}  \Bv_{\CR} \Bh_{k}  \Sfc \left( W^{\left< k \right>_{\CR}}_{\CR \backslash \left\lbrace  \left< k \right>_{\CR } \right\rbrace  ,q} \right)  + \Bv_{\CR} \Bn_{\rm BS} ,
    \end{aligned}
\end{equation}
which  is a codeword available in the set $\CB$ defined in~\eqref{eq: set B ref}. 

Accordingly, during stage 1,  as per~\eqref{eq: codeword UL set A} and \eqref{eq: Ysl code final}, the BS can create all codewords in the set $\CA \cup \CB$.
Hence, the total number of created codewords during stage 1 is given by:
\begin{equation}
\label{eq: sum sets A B}
    \begin{aligned}
        \left\vert \CA \right\vert + \left\vert \CB \right\vert= \binom{t+L-1}{t}+\binom{t+L-1}{t+1}=\binom{t+L}{t+1}.
    \end{aligned}
\end{equation}
Similarly, for each stage $i \in [N_{\rm S}]$, by following the same way as in~\eqref{eq: codeword UL set A}-\eqref{eq: Ysl code final}, one can simply show that the BS is able to create all codewords of size $t+1$, each containing the subpackets of users $k \in \CU (i)$. The BS uses these created codewords in the DL step. 


Now, in order to compute the NDT for the UL step, we set $N_{\rm UL} = N_{\rm S} N_{\rm T}$. Next, by applying \eqref{eq: T UL def.}, the NDT for the UL step is given by:

\begin{equation}
    \begin{aligned}
        T_{\rm UL} = N_{\rm UL}  Q = \frac{N_{\rm S} N_{\rm T}}{\binom{K}{t} \binom{K-t-1}{L-1}} = \frac{K-t}{t+L}.
    \end{aligned}
\end{equation}

\section{Proof of Theorem~\ref{theorem: DL}}
\label{proof: DL}
 Similarly to the UL step in Section~\ref{section: UL A}, we first focus on stage 1, where users $\CU (1)=[t+L]$ are served by the BS, and show that all users in the set $\CU (1)$ are able to decode all subpackets of their requested files. Then, one can simply extend the process to all stages $i \in \left[ N_{\rm S} \right]$. To this end, consider the $j$-th transmission of stage 1. According to~\eqref{eq: y k system def} and \eqref{eq: XBS j A}, during the transmission $j \in \left[ N_{\rm T} \right]$ of stage 1, user $k \in \left[ t+L \right]$, receives the signal
\begin{equation}
\label{eq: y k A}
    \begin{aligned}
        y_{k} (j) & = \Bh_{k}\herm \Bx_{\mathrm{BS}} (j) + n_{k} \\
        & = \sum\limits_{\CQ \subseteq [t+L], \left\vert \CQ \right\vert =t+1 }  \Ba_{\CQ} [j] \Bh_{k}\herm \Bw_{\CQ} f_{\CQ} + n_{k}.
    \end{aligned} 
\end{equation}

 Then, by applying the ZF precoders into  \eqref{eq: y k A}, user $k \in [t+L]$ observes the signal:
 \begin{equation}
 \label{eq: y K A ref}
     \begin{aligned}
         y_{k} (j) &= \sum\limits_{ \CQ \in \CQ_{k}}  \Ba_{\CQ} [j] \Bh_{k}\herm \Bw_{\CQ} f_{\CQ} + n_{k},
     \end{aligned}
 \end{equation}
 where $\CQ_{k}=\left\lbrace \CQ: \CQ \subseteq \left[ t+L \right], \left\vert \CQ \right\vert=t+1, k \in \CQ \right\rbrace$. As per~\eqref{eq: XBS j A}, recall that $f_{\CQ}$ is expressed as 
 \begin{equation}
\label{eq: f k Q def}
     \begin{aligned}
         f_{\CQ}=  \sum\limits_{l \in \CQ} \theta_{\CQ} \Bv_{\CQ} \Bh_{l} \Sfc \big( W^{\left< l \right>_{\CQ}}_{\CQ \backslash \lbrace  \left< l \right>_{\CQ} \rbrace ,q} \big) +  \Bv_{\CQ} \Bn_{\rm BS},
     \end{aligned}
 \end{equation}
 such that 
 \begin{equation}
     \begin{aligned}
         \theta_{\CQ} =
         \begin{cases}
             1  & 1 \in \CQ \\
             -1 &  1 \notin \CQ
         \end{cases}.
     \end{aligned}
 \end{equation}
 
 Now, let us focus on the decoding process at user $k \in [t+L]$. To this end, for $\CQ \in \CQ_{k}$, and $l^{*} \in \CQ$, assume that $\left< l^{*} \right>_{\CQ}=k$. In other words, the element $l^{*} \in \CQ$ is the next element to $k \in \CQ$ according to a circular shift on $\CQ$ (cf. \eqref{eq: shift def}). 
 Therefore, we can rewrite \eqref{eq: f k Q def} as follows
 \begin{equation}
 \label{eq: f Q modified}
    \begin{aligned}
        f_{\CQ} & = \sum\limits_{l \in \CQ \backslash \lbrace l^{*} \rbrace} \theta_{\CQ} \Bv_{\CQ} \Bh_{l} \Sfc \left( W^{\left< l \right>_{\CQ}}_{\CQ \backslash \left\lbrace  \left< l \right>_{\CQ} \right\rbrace ,q} \right) \\
        & + \theta_{\CQ} \Bv_{\CQ} \Bh_{l^{*}} \Sfc \left( W^{k}_{\CQ \backslash \left\lbrace  k \right\rbrace ,q} \right) +  \Bv_{\CQ} \Bn_{\rm BS}.
    \end{aligned}
\end{equation}
Here, since $\CQ \in \CQ_{k}$ and $k \in \CQ \backslash \lbrace l^{*} \rbrace$, user $k$ has all subpackets $ W^{\left< l \right>_{\CQ}}_{\CQ \backslash \left\lbrace  \left< l \right>_{\CQ} \right\rbrace ,q}$ in its cache memory for all $l \in \CQ \backslash \lbrace l^{*} \rbrace$. Therefore, it can regenerate the term $\sum\nolimits_{l \in \CQ \backslash \lbrace l^{*} \rbrace} \theta_{\CQ}  \Bv_{\CQ} \Bh_{l} \Sfc \big( W^{\left< l \right>_{\CQ}}_{\CQ \backslash \lbrace  \left< l \right>_{\CQ} \rbrace ,q} \big)$ and subtract it from \eqref{eq: f Q modified}, to observe the signal $\hat{f}_{\CQ}$ as follows:
\begin{equation}
\label{eq: f K Q final}
    \begin{aligned}
        \hat{f}_{\CQ} =  \theta_{\CQ}  \Bv_{\CQ} \Bh_{l^{*}} \Sfc \left( W^{k}_{\CQ \backslash \left\lbrace  k \right\rbrace ,q} \right)
        + \Bv_{\CQ} \Bn_{\rm BS} .
    \end{aligned}
\end{equation}
Therefore, for $j \in [N_{\rm T}]$, by applying~\eqref{eq: f K Q final} into~\eqref{eq: y K A ref}, $y_{k} (j)$ is simplified to $\hat{y}_{k} (j)$, which is expressed as:
 \begin{equation}
 \label{eq: y hat def}
     \begin{aligned}
         \hat{y}_{k} (j) &= \sum\limits_{ \CQ \in \CQ_{k}}  \Ba_{\CQ} [j] \Bh_{k}\herm \Bw_{\CQ} \hat{f}_{\CQ} + n_{k}.
     \end{aligned}
 \end{equation}

 Here, according to that $\left\vert \CQ_{k} \right\vert = N_{\rm T}=\binom{t+L-1}{t}$ and $j \in \left[ N_{\rm T} \right]$, we have a system of equations with $N_{\rm T}$ unknowns and $N_{\rm T}$ equations to solve $\hat{f}_{\CQ}$. In order to solve this system of equations, first, define $\hat{\By}_{k}=\left[ \hat{y}_{k} (1), \cdots, \hat{y}_{k} (N_{\rm T} ) \right]^{\rm T}$,  and $\Bn_{k}=\left[ n_{k}, \cdots, n_{k} \right]^{\rm T}$ with $ \Bn_{k} \in \mathbb{C}^{N_{\rm T} \times 1}$. Moreover, let
 $\BA_{\CQ_{k}} = \left[ \Ba_{\CQ_{k}(1)}^{\rm T}, \cdots, \Ba_{\CQ_{k} \left( N_{\rm T} \right)}^{\rm T}  \right]$, where $Q_{k}(i)$ represents the $i$-th entry of $\CQ_{k}$. As per Section~\ref{section: DL step}, we recall that for $\CQ \subseteq [t+L]$ with $\left\vert \CQ \right\vert=t+1$, the BS creates a random vector $\Ba_{\CQ} \in \mathbb{R}^{1 \times N_{\rm T}}$. As a result, $\Ba_{\CQ_{k}(i)}$ demonstrates the random vector generated by the BS for the set $\CQ_{k}(i)$. 
 Hence, by using \eqref{eq: y hat def}, the system of equations takes the form as follows:
 \begin{equation}
 \label{eq: y K A sys eq}
     \begin{aligned}
         \hat{\By}_{k} = \BA_{\CQ_{k}}
         \begin{bmatrix}
             \Bh_{k}\herm \Bw_{\CQ_{k}(1)} \hat{f}_{\CQ_{k}(1)} \\
             \vdots \\
            \Bh_{k}\herm \Bw_{\CQ_{k}(N_{\rm T})} \hat{f}_{\CQ_{k}(N_{\rm T})} \\
         \end{bmatrix}
         + \Bn_{k}.
     \end{aligned}
 \end{equation}
Now, by utilizing \eqref{eq: f K Q final} and \eqref{eq: y K A sys eq}, and setting $\Tilde{\By}_{k}=\BA_{\CQ_{k}}^{-1} \hat{\By}_{k}$ and $\Tilde{\Bn}_{k}=\BA_{\CQ_{k}}^{-1} \Bn_{k} $, for each $j \in [N_{\rm T}]$, it is observed that:
\begin{equation}
\label{eq: f k Q j A}
    \begin{aligned}
       \Tilde{\By}_{k}[j] & \! = \! \Bh_{k}\herm \Bw_{\CQ_{k}(j)} \hat{f}_{\CQ_{k}(j)} + \Tilde{\Bn}_{k}[j] \\
       & =  \Bh_{k}\herm \Bw_{\CQ_{k}(j)} \theta_{\CQ_{k}(j)} \Bv_{\CQ_{k}(j)} \Bh_{l^{*}} \Sfc \left( W^{k}_{\CQ_{k}(j) \backslash \left\lbrace  k \right\rbrace ,q} \right) \\
       & + \Bh_{k}\herm \Bw_{\CQ_{k}(j)} \Bv_{\CQ_{k}(j)} \Bn_{\rm BS} + \Tilde{\Bn}_{k}[j],
    \end{aligned}
\end{equation}
where $\Tilde{\By}_{k}[j]$ and $\Tilde{\Bn}_{k}[j]$ represent the $j$-th entry of $\Tilde{\By}_{k}$ and $\Tilde{\Bn}_{k}$, respectively. As a result, by employing \eqref{eq: f k Q j A}, for each $\CQ \subseteq [t+L]$ with $\left\vert \CQ \right\vert =t+1$ and $k \in \CQ$, user $k$ can recover $\Sfc \big( W^{k}_{\CQ \backslash \lbrace  k \rbrace ,q} \big)$. Next, user $k$ decodes and estimates $\Sfc \big( W^{k}_{\CQ \backslash \lbrace  k \rbrace ,q} \big)$ to retrieve the subpacket $\hat{W}^{k}_{\CQ \backslash \lbrace  k \rbrace ,q}$.

Now, according to the fact that for any $k \in [t+L]$, we have $\left\lvert \CQ_{k} \right\rvert = N_{\rm T}$, and for each $\CQ \in \CQ_{k}$, user $k$ is able to recover the subpacket $\hat{W}^{k}_{\CQ \backslash \lbrace  k \rbrace ,q}$, the total number of subpackets retrieved by user $k$ during stage 1 is $N_{\rm T}$. Similarly, by changing the users' indices in \eqref{eq: y k A}-\eqref{eq: f k Q j A}, one can simply show that if user $k \in [K]$ is served during stage $j \in [N_{\rm S}]$, then it can decode $N_{\rm T}$ subpackets of its requested file in this stage. Here, since user $k \in [K]$ is served in $\binom{K-1}{t+L-1}$ stages of the DL step, it is able to decode $N_{\rm T} \binom{K-1}{t+L-1} $ of its requested subpackets during the DL step. On the other hand, user $k$ has stored $\gamma \binom{K}{t} \binom{K-t-1}{L-1}$ subpackets of its desired file before starting the UL-DL steps. Hence, after the UL-DL communication phase, by using $t=K\gamma$, the total number of available useful subpackets at user $k$, i.e., the subpackets for its requested file, takes the form as follows:
\begin{equation}
    \begin{aligned}
        & N_{\rm T} \binom{K-1}{t+L-1}  +  \gamma \binom{K}{t} \binom{K-t-1}{L-1}  = \\
        & \binom{K}{t} \binom{K-t-1}{L-1},
    \end{aligned}
\end{equation}
which is equal to the total number of subpackets for the file $W^{k}$, and shows that user $k \in [K]$ is able to decode its requested file after the DL step. 

In order to compute the achievable NDT for the DL step, we follow the same way as Appendix~\ref{proof: theorem  UL}. In this regard, by substituting $N_{\rm DL} = N_{\rm S} N_{\rm T}$ and $Q=\frac{1}{\binom{K}{t} \binom{K-t-1}{L-1}}$ into~\eqref{eq: T DL def.}, the NDT for the DL step is computed as follows
\begin{equation}
       \begin{aligned}
           T_{\rm DL} = N_{\rm DL} Q = \frac{N_{\rm S} N_{\rm T}}{\binom{K}{t} \binom{K-t-1}{L-1}} = \frac{K-t}{t+L}.
       \end{aligned}
   \end{equation}

\section{An Example for $K=5$, $t=2$ and $L=3$}
\label{exmp: general}
 Consider a cache-aided MISO network with $L=3$, $\gamma=0.4$, $K=5$ and $t=2$, such that users $1$, $2$, $3$, $4$ and $5$ request the files $W^{1}$, $W^{2}$, $W^{3}$, $W^{4}$ and $W^{5}$, respectively. First, each file $W^{k}$, $k \in [5]$, is split into $\binom{K}{t}=10$ packets $W^{k}_{\CP}$, where $\CP \subset [5]$ and $\left\vert \CP \right\vert=t=2$. Then, for each $\CP \subset [5]$ with $\left\vert \CP \right\vert=2$, user $k \in [\CP]$ stores the packet $W^{n}_{\CP}$ for $n \in [5]$. 
 
 Prior to the UL-DL steps, each packet $W^{n}_{\CP}$ is split into $q=\binom{K-t-1}{L-1}=1$ subpacket $W^{n}_{\CP,q}$. Hereafter, in order to simplify the notation, we drop the index $q$ in $W^{n}_{\CP,q}$, and represent the subpacket $W^{n}_{\CP,q}$ by $W^{n}_{\CP}$. 

    \textbf{1) UL Step: } The UL step consists of $N_{\rm S}=\binom{K}{t+L}=1$ stage, and this stage involves $N_{\rm T}= \binom{t+L-1}{t}=6$ transmissions. In order to represent the users' transmitted signals in each transmission, by using \eqref{eq: set M def}, we define the set $\CM$ as follows: 
    \begin{equation}
    \CM=\left\lbrace \CS: \CS \subset \left[ 5 \right],  1  \in \CS, \left\vert \CS \right\vert=3 \right\rbrace,
\end{equation}
where $\left\vert \CM \right\vert=N_{\rm T}$, and each $\CS \in \CM$ shows an index for a separate transmission. For example, for $\CS=\lbrace 123 \rbrace$, $x^{k} ({\lbrace 123 \rbrace})$ shows the transmitted signal of user $k$ in the transmission associated with the set $\lbrace 1 2 3 \rbrace$. 

Next, for each $\CP \subset [5]$ with $\left\vert \CP \right\vert =2$, each user encodes its cached subpackets, such that $\Sfc \left( W^{1}_{\CP} \right)=A_{\CP}$, $\Sfc \left( W^{2}_{\CP} \right)=B_{\CP}$, $\Sfc \left( W^{3}_{\CP} \right)=C_{\CP}$, $\Sfc \left( W^{4}_{\CP} \right)=D_{\CP}$ and $\Sfc \left( W^{5}_{\CP} \right)=E_{\CP}$. Moreover, for $\CQ \subset [5]$ with $\left\lvert \CQ \right\vert = 3$, and $k \in [5]$, it is assumed that
\begin{equation}
\label{eq: power allocation exmp A}
    \begin{aligned}
        \mathbb{E} \left[ \left\vert  \Sfc \left( W^{\left< k \right>_{\CQ}}_{\CQ \backslash \left\lbrace \left< k \right>_{\CQ} \right\rbrace } \right) \right\vert^{2} \right]=
    \end{aligned}
    \begin{cases}
       P_{\rm UL}  & k=1 \\
        \frac{P_{\rm UL}}{2} & k \neq 1
    \end{cases},
\end{equation}
where $P_{\rm UL}$ represents the transmit power of user $1$ during the transmission $\CS \in \CM$.
 As observed, the energy consumption of user $k \in [5]$ during the UL step is equal to $E_{k} = P_{\rm UL} N_{\rm T} \binom{K-1}{t+L-1}=6P_{\rm UL}$.

Now, by utilizing \eqref{eq: x KS A def}, during the transmission $\CS \in \CM$, user $k \in [5]$ transmits the signal $x^{k} (\CS)$ shown in Table~\ref{table: UL TX A}. In order to give further insight into Table~\ref{table: UL TX A} and \eqref{eq: x KS A def}, let us focus on the transmission $\lbrace  123 \rbrace$. Here,  user $k \in \lbrace 1,2,3 \rbrace$ transmits the encoded subpacket of the requested file of user $\left< k \right>_{\lbrace 123 \rbrace }$, which is stored at user $k$ and not available at user $\left< k \right>_{\lbrace 123 \rbrace}$, i.e., $x^{k} ( \lbrace 123 \rbrace ) = W^{\left< k \right>_{\lbrace 123 \rbrace}}_{\lbrace 123 \rbrace \backslash \lbrace \left< k \right>_{\lbrace 123 \rbrace} \rbrace} $. For instance, user $1$ transmits the requested subpacket of user $\left< 1 \right>_{\{123 \}}=2$, which is $x^{1} ( \lbrace 123 \rbrace ) = B_{\lbrace 1 2 3 \rbrace \backslash \left< 1 \right>_{\lbrace 123 \rbrace}}=B_{13}$. Moreover, during the transmission $\lbrace 123 \rbrace $, user $k \in \lbrace 4,5 \rbrace$ transmits a superposition of the encoded subpackets of the files requested by users $\left\lbrace \left< k \right>_{\lbrace 123 \rbrace \cup \lbrace k \rbrace \backslash \lbrace j \rbrace} : j \in [3] \right\rbrace$ that are stored at user $k$, i.e.,
\begin{equation*}
    x^{k} (\lbrace 123 \rbrace)= - \sum\limits_{j \in \lbrace 1,2,3 \rbrace}   \Sfc \left( W^{ \left< k \right>_{\lbrace 123 \rbrace \cup \left\lbrace k \right\rbrace  \backslash \left\lbrace j \right\rbrace} }_{ \lbrace 123 \rbrace \cup \left\lbrace k \right\rbrace \backslash \left\lbrace j, \left< k \right>_{\lbrace 123 \rbrace \cup \left\lbrace k \right\rbrace \backslash \left\lbrace j \right\rbrace } \right\rbrace } \right).
\end{equation*}
Accordingly, user $4$, for example, transmits the signal $x^{4} (\lbrace 1 2 3 \rbrace ) = -B_{34}- A_{34}-A_{24}$.

\begin{table}[t]
\Large
  \centering
  \caption{Users' Transmitted Signals During the UL Step}
  \resizebox{\columnwidth}{!}{
  \begin{tabular}{|c|c|c|c|c|c|}
    \hline
    \multicolumn{1}{|c|}{{\diagbox[innerwidth=1.9cm]{$\CS$}{$x^{k} (\CS)$}}} & $x^{1} (\CS)$ & $x^{2} (\CS)$ &  $x^{3} (\CS)$ &  $x^{4} (\CS)$ & $x^{5} (\CS)$ \\
    \cline{3-6}
    \hline
    \multirow{2}{*}{$\lbrace 123\rbrace$} &  \multirow{2}{*}{$B_{13}$} &  \multirow{2}{*}{$C_{12}$} &  \multirow{2}{*}{$A_{23}$} & 
    \multirow{2}{*}{\begin{tabular}{@{}c@{}}$-B_{34}$ \\$- A_{34} -  A_{24} $\end{tabular}}    & \multirow{2}{*}{\begin{tabular}{@{}c@{}}$-B_{35}$ \\$ -  A_{35} -  A_{25} $\end{tabular}} \\
    &&&&& \\
    \hline
\multirow{2}{*}{$\lbrace 124\rbrace$} & \multirow{2}{*}{$B_{14}$} & \multirow{2}{*}{$D_{12}$} &  \multirow{2}{*}{\begin{tabular}{@{}c@{}}$-D_{23}$ \\$- D_{13} -  A_{23} $\end{tabular}} & \multirow{2}{*}{$A_{24}$} & \multirow{2}{*}{\begin{tabular}{@{}c@{}}$-B_{45}$ \\$-   A_{45} -  A_{25} $\end{tabular}} \\
 &&&&& \\
    \hline
    \multirow{2}{*}{$\lbrace 125\rbrace$} & \multirow{2}{*}{$B_{15}$} & \multirow{2}{*}{$E_{12}$} &  \multirow{2}{*}{\begin{tabular}{@{}c@{}}$-E_{23}$ \\$- E_{13} -  A_{23} $\end{tabular}}  & \multirow{2}{*}{\begin{tabular}{@{}c@{}}$-E_{24}$ \\$- E_{14} -  A_{24} $\end{tabular}}  &  \multirow{2}{*}{$A_{25}$} \\
     &&&&& \\
    \hline
      \multirow{2}{*}{$\lbrace 134\rbrace$} &  \multirow{2}{*}{$C_{14}$} &  \multirow{2}{*}{\begin{tabular}{@{}c@{}}$-C_{24}$ \\$- D_{12} -  C_{12} $\end{tabular}}  &  \multirow{2}{*}{$D_{13}$} &  \multirow{2}{*}{$A_{34}$} & \multirow{2}{*}{\begin{tabular}{@{}c@{}}$-C_{45}$ \\$- A_{45} -  A_{35} $\end{tabular}} \\
      &&&&& \\
    \hline
     \multirow{2}{*}{$\lbrace 135\rbrace$} & \multirow{2}{*}{$C_{15}$} & \multirow{2}{*}{\begin{tabular}{@{}c@{}}$-C_{25}$ \\$-  E_{12} -  C_{12} $\end{tabular}} & \multirow{2}{*}{$E_{13}$} & \multirow{2}{*}{\begin{tabular}{@{}c@{}}$-E_{34}$ \\$- E_{14} -  A_{34} $\end{tabular}}  & \multirow{2}{*}{$A_{35}$} \\
       &&&&& \\
    \hline
     \multirow{2}{*}{$\lbrace 145\rbrace$} & \multirow{2}{*}{$D_{15}$} & \multirow{2}{*}{\begin{tabular}{@{}c@{}}$-D_{25}$ \\$- E_{12} -  D_{12}  $\end{tabular}} &  \multirow{2}{*}{\begin{tabular}{@{}c@{}}$-D_{35}$ \\$- E_{13} -  D_{13} $\end{tabular}}  & \multirow{2}{*}{$E_{14}$} & \multirow{2}{*}{$A_{45}$} \\
      &&&&& \\
    \hline
  \end{tabular}
  }
  \label{table: UL TX A}
  \end{table}

Now, by using \eqref{eq: yBS A}, during the transmission $\CS \in \CM$, the BS receives the signal
\begin{equation}
\label{eq: yBS exmp A}
    \begin{aligned}
        \By_{\rm BS} (\CS)=\sum\limits_{k \in \left[ 5 \right]} \Bh_{k} x^{k} ( \CS ) +\Bn_{\rm BS}.
    \end{aligned}
\end{equation}

Following $N_{\rm T}$ UL transmissions, the BS creates two sets of codewords, denoted as $\CA$ and $\CB$, which are illustrated in \eqref{eq: set A def} and \eqref{eq: set B ref}, respectively. Generally speaking, each codeword in $\CA$ contains $t+1$ subpackets for users in $\CS \subset [5]$ with $\left\vert \CS \right\vert =t+1$ and $1 \in \CS $, while each codeword of $\CB$ is composed of $t+1$ subpackets for users in $\CR \subset [5]$ with $\left\vert \CR \right\vert =t+1$ and $1 \notin \CR$. In order to create the codewords in the set $\CA$, for each $\CS \subset [5] $ with $\left\vert \CS \right\vert=t+1$ and $1 \in \CS$, we multiply \eqref{eq: yBS exmp A} by beamforming (row) vector $\Bv_{\CS} \in \mathbb{C}^{1 \times 3} $ as follows
\begin{equation}
    \begin{aligned}
        \Bv_{\CS}\By_{\rm BS} (\CS ) = \sum\limits_{k \in \CS} \Bv_{\CS} \Bh_{k} x^{k} ( \CS ) +\Bv_{\CS}\Bn_{\rm BS}, 
    \end{aligned}
\end{equation}
which represents the codeword $\sum\limits_{k \in \CS} \Bv_{\CS} \Bh_{k} x^{k} ( \CS )$ added up by the noise $\Bv_{\CS}\Bn_{\rm BS}$. For instance, for $\CS=\lbrace 123 \rbrace$ and $\CS=\lbrace 135 \rbrace$, the BS generates the codewords 
\begin{equation*}
    \begin{aligned}
        \Bv_{\lbrace 123 \rbrace}\By_{\rm BS} (\lbrace 123 \rbrace) & = \Bv_{\lbrace 123 \rbrace} ( \Bh_{1} B_{13}  + \Bh_{2} C_{12}  + \Bh_{3} A_{23} )  \\
        & +\Bv_{\lbrace 123 \rbrace}\Bn_{\rm BS}, \\
          \Bv_{\lbrace 135 \rbrace}\By_{\rm BS} (\lbrace 135 \rbrace) & = \Bv_{\lbrace 135 \rbrace} ( \Bh_{1} C_{15}    + \Bh_{3} E_{13}   + \Bh_{5} A_{35} )  \\
          & +\Bv_{\lbrace 135 \rbrace}\Bn_{\rm BS}.
    \end{aligned}
\end{equation*}

Moreover, to generate the codewords in the set $\CB$, for each $\CR \subset [5]$, with $\left\vert \CR \right\vert=t+1$ and $1 \notin \CR$, by following the same way as in~\eqref{eq: Ysl code def}-\eqref{eq: Ysl code final}, we compute $ \sum\nolimits_{j \in \CR }  \Bv_{\CR} \By_{\rm BS} \left( \CR \cup \lbrace 1 \rbrace \backslash \lbrace j \rbrace \right)$ as follows
  \begin{equation*}
    \begin{aligned}
        \sum\limits_{j \in \CR }  \Bv_{\CR} \By_{\rm BS} \left( \CR \cup \lbrace 1 \rbrace \backslash \lbrace j \rbrace \right) & = 
        - \sum\limits_{k \in \CR}  \Bv_{\CR } \Bh_{k} \Sfc \left( W^{\left< k \right>_{\CR}}_{\CR  \backslash \left\lbrace  \left< k \right>_{\CR } \right\rbrace } \right) \\
&        + \Bv_{\CR} \Bn_{\rm BS}.
    \end{aligned}
\end{equation*}
For example, for $\CR=\lbrace 234 \rbrace$ and $\CR=\lbrace 345 \rbrace$, the BS creates the following codewords
\begin{equation*}
    \begin{aligned}
          \sum\limits_{j \in \lbrace 234 \rbrace }  &\Bv_{\lbrace 234 \rbrace} \By_{\rm BS} \left( \lbrace 234 \rbrace \cup \lbrace 1 \rbrace \backslash \lbrace j \rbrace \right) = \Bv_{\lbrace 234 \rbrace} \By_{\rm BS} (\lbrace 134 \rbrace ) + \\ 
          & \Bv_{\lbrace 234 \rbrace} \By_{\rm BS} (\lbrace 124 \rbrace ) + \Bv_{\lbrace 234 \rbrace} \By_{\rm BS} (\lbrace 123 \rbrace ) = \\
        - &  \Bv_{\lbrace 234 \rbrace} \left( \Bh_{2} C_{24} + \Bh_{3} D_{23} + \Bh_{4} B_{34} \right) + \Bv_{\lbrace 234 \rbrace}\Bn_{\rm BS},
    \end{aligned}
\end{equation*}
\begin{equation*}
    \begin{aligned}
          \sum\limits_{j \in \lbrace 345 \rbrace }  &\Bv_{\lbrace 345 \rbrace} \By_{\rm BS} \left( \lbrace 345 \rbrace \cup \lbrace 1 \rbrace \backslash \lbrace j \rbrace \right) = \Bv_{\lbrace 345 \rbrace}  \By_{\rm BS} \left( \lbrace 145 \rbrace \right) + \\
         & \Bv_{\lbrace 345 \rbrace}  \By_{\rm BS} \left( \lbrace 135 \rbrace \right) + \Bv_{\lbrace 345 \rbrace}  \By_{\rm BS} \left( \lbrace 134 \rbrace \right) = \\
        - &  \Bv_{\lbrace 345 \rbrace} \left( \Bh_{3} D_{35} + \Bh_{4} E_{34} + \Bh_{5} C_{45} \right) + \Bv_{\lbrace 345 \rbrace}\Bn_{\rm BS}.
    \end{aligned}
\end{equation*}
As mentioned, during the UL step, there exist $N_{\rm UL} = N_{\rm S} N_{\rm T} = 6$ transmissions, and $Q = \frac{1}{\binom{K}{t} \binom{K-t-1}{L-1}}=\frac{1}{10}$. Hence, as per~\eqref{eq: T UL def.} and Theorem~\ref{theorem: load UL A}, the NDT for the UL step is given by:
\begin{equation}
     T_{\rm UL} = \frac{N_{\rm UL} }{\binom{K}{t} \binom{K-t-1}{L-1}} = \frac{K-t}{t+L}=\frac{3}{5}.
\end{equation}

\textbf{2) DL Step: } The DL step is also comprised of $N_{\rm S}=1$ stage with $N_{\rm T}=6$ transmissions. During each transmission, the BS transmits a random superposition of the generated codewords in the sets $\CA$ and $\CB$. To this end, for each $\CQ \subset [5]$ with $\left\vert \CQ \right\vert=t+1$, the BS generate a random vector $\Ba_{\CQ} \in \mathbb{R}^{1 \times N_{\rm T}}$ and transmits the following signal during the transmission $j \in [N_{\rm T}]$.
\begin{equation}
    \begin{aligned}
        & \Bx_{\mathrm{BS}} (j) = \sum\limits_{\substack{\CQ  \subset [5],  \left\vert \CQ \right\vert=t+1}}  \Ba_{\CQ} [j]  \Bw_{\CQ} f_{\CQ},
    \end{aligned}
\end{equation}
where $ f_{\CQ}=\Bv_{\CQ} \Bg_{\CQ} +\Bv_{\CQ} \Bn_{\rm BS}$, and the vectors $\Bg_{\CQ}$ are defined in Table~\ref{table: gQ}. Here, we note that for each $\CQ \subset [5]$ with $\left\vert \CQ \right\vert=t+1$, each $f_{\CQ}$ denotes a noisy codeword of size $t+1$ that is created during the UL step.
\begin{table}[htb]
    \centering
    \large
     \caption{The Vectors $\Bg_{\CQ}$ for Different Sets of $\CQ$}
      \resizebox{\columnwidth}{!}{
    \begin{tabular}{|l|l|}
        \hline
        $\Bg_{\lbrace 123 \rbrace} \! \! = \! \Bh_{1} B_{13} \! + \!  \Bh_{2} C_{12} \! + \! \Bh_{3} A_{23}$ &  $\Bg_{\lbrace 124 \rbrace} \! \!= \! \Bh_{1} B_{14} \! + \!  \Bh_{2} D_{12} \! + \!  \Bh_{4} A_{24}$ \\
        \hline
        $\Bg_{\lbrace 125 \rbrace} \! \! = \! \Bh_{1} B_{15} \! + \!  \Bh_{2} E_{12} \! + \! \Bh_{5} A_{25}$ &  $\Bg_{\lbrace 134 \rbrace} \! \!= \! \Bh_{1} C_{14} \! + \!  \Bh_{3} D_{13} \! + \!  \Bh_{4} A_{34}$ \\
        \hline
        $\Bg_{\lbrace 135 \rbrace} \! \! = \! \Bh_{1} C_{15} \! + \!  \Bh_{3} E_{13} \! + \! \Bh_{5} A_{35}$ &  $\Bg_{\lbrace 145 \rbrace} \! \!= \! \Bh_{1} D_{15} \! + \!  \Bh_{4} E_{14} \! + \!  \Bh_{5} A_{45}$ \\
        \hline
        $\Bg_{\lbrace 234 \rbrace} \! \! = \! -\Bh_{2} C_{24} \! - \!  \Bh_{3} D_{23} \! - \! \Bh_{4} B_{34}$ &  $\Bg_{\lbrace 235 \rbrace} \! \!= \! - \Bh_{2} C_{25} \! - \!  \Bh_{3} E_{23} \! - \!  \Bh_{5} B_{35}$ \\
        \hline
        $\Bg_{\lbrace 245 \rbrace} \! \! = \! - \Bh_{2} D_{25} \! - \!  \Bh_{4} E_{24} \! - \! \Bh_{5} B_{45}$ &  $\Bg_{\lbrace 345 \rbrace} \! \!= \! - \Bh_{3} D_{35} \! - \!  \Bh_{4} E_{34} \! - \!  \Bh_{5} C_{45}$ \\
        \hline
    \end{tabular}
    }
    \label{table: gQ}
\end{table}

In the proceeding, we demonstrate that each user is able to decode all of its requested subpackets. In this regard, without loss of generality (up to the permutation of users' indices), consider user $1$. Hence, during the DL transmission $j \in [N_{\rm T}],$ by using \eqref{eq: y K A ref} and applying the ZF precoders, user $1$ receives the signal 
\begin{equation}
\label{eq: y 1 j g exmp}
     \begin{aligned}
         y_{1} (j) &=  \Ba_{\lbrace 123 \rbrace} [j] \Bh_{1}\herm \Bw_{\lbrace 123 \rbrace} f_{\lbrace 123 \rbrace} \! + \!
          \Ba_{\lbrace 124 \rbrace} [j] \Bh_{1}\herm \Bw_{\lbrace 124\rbrace} f_{\lbrace 124 \rbrace}\\
        &+  \Ba_{\lbrace 125 \rbrace} [j] \Bh_{1}\herm \Bw_{\lbrace 125 \rbrace} f_{\lbrace 125 \rbrace} \! + \!
         \Ba_{\lbrace 134 \rbrace} [j] \Bh_{1}\herm \Bw_{\lbrace 134 \rbrace} f_{\lbrace 134 \rbrace} \\
         &+  \Ba_{\lbrace 135 \rbrace} [j] \Bh_{1}\herm \Bw_{\lbrace 135 \rbrace} f_{\lbrace 135 \rbrace} 
       \!  + \! \Ba_{\lbrace 145 \rbrace} [j] \Bh_{1}\herm \Bw_{\lbrace 145 \rbrace} f_{\lbrace 145 \rbrace}\\
         & + n_{1}.
     \end{aligned}
 \end{equation}
User 1 has stored $B_{13}$, $C_{12}$, $B_{14}$, $D_{12}$, $B_{15}$, $E_{12}$ $C_{14}$, $D_{13}$, $C_{15}$, $E_{13}$, $D_{15}$ and $E_{14}$ in its cache memory. Hence, by utilizing Table~\ref{table: gQ} and following the same way as in~\eqref{eq: f k Q def}-\eqref{eq: y hat def}, we remove the interference of the cached contents from~\eqref{eq: y 1 j g exmp}, which simplifies $y_{1}(j)$ to $\hat{y}_{1} (j)$ as follows
\begin{equation}
\label{eq: y 1 hat j g exmp}
     \begin{aligned}
         y_{1} (j) &= \Bb_{ \lbrace 23 \rbrace} [j] A_{23}  + \Ba_{\lbrace 123 \rbrace} [j] \Bh_{1}\herm \Bw_{\lbrace 123 \rbrace} \Tilde{n}_{\lbrace 123 \rbrace} \\
         & + \Bb_{ \lbrace 24 \rbrace} [j] A_{24}  +  \Ba_{\lbrace 124 \rbrace} [j] \Bh_{1}\herm \Bw_{\lbrace 124 \rbrace} \Tilde{n}_{\lbrace 124 \rbrace} \\
         & + \Bb_{ \lbrace 25 \rbrace} [j] A_{25}  +  \Ba_{\lbrace 125 \rbrace} [j] \Bh_{1}\herm \Bw_{\lbrace 125 \rbrace} \Tilde{n}_{\lbrace 125 \rbrace} \\
         & + \Bb_{ \lbrace 34 \rbrace} [j] A_{34}  +  \Ba_{\lbrace 134 \rbrace} [j] \Bh_{1}\herm \Bw_{\lbrace 134 \rbrace} \Tilde{n}_{\lbrace 134 \rbrace} \\
         & + \Bb_{ \lbrace 35 \rbrace} [j] A_{35}  +  \Ba_{\lbrace 135 \rbrace} [j] \Bh_{1}\herm \Bw_{\lbrace 135 \rbrace} \Tilde{n}_{\lbrace 135 \rbrace} \\
         & + \Bb_{ \lbrace 45 \rbrace} [j] A_{45}  +  \Ba_{\lbrace 145 \rbrace} [j] \Bh_{1}\herm \Bw_{\lbrace 145 \rbrace} \Tilde{n}_{\lbrace 145 \rbrace} + n_{1},
     \end{aligned}
 \end{equation}
where $\Tilde{n}_{\CQ}= \Bv_{\CQ} \Bn_{\rm BS}$ and
\begin{equation}
    \begin{aligned}
       & \Bb_{ \lbrace 23 \rbrace} [j] =\Ba_{\lbrace 123 \rbrace} [j] \Bv_{\lbrace 123 \rbrace} \Bh_{3} \Bh_{1}\herm \Bw_{\lbrace 123 \rbrace} \\
       & \Bb_{ \lbrace 24 \rbrace} [j]= \Ba_{\lbrace 124 \rbrace} [j] \Bv_{\lbrace 124 \rbrace} \Bh_{4} \Bh_{1}\herm \Bw_{\lbrace 124 \rbrace}  \\
       & \Bb_{ \lbrace 25 \rbrace} [j]= \Ba_{\lbrace 125 \rbrace} [j] \Bv_{\lbrace 125 \rbrace} \Bh_{5} \Bh_{1}\herm \Bw_{\lbrace 125 \rbrace}  \\
        & \Bb_{ \lbrace 34 \rbrace} [j]=  \Ba_{\lbrace 134 \rbrace} [j] \Bv_{\lbrace 134 \rbrace} \Bh_{4} \Bh_{1}\herm \Bw_{\lbrace 134 \rbrace}  \\
        & \Bb_{ \lbrace 35 \rbrace} [j]=   \Ba_{\lbrace 135 \rbrace} [j] \Bv_{\lbrace 135 \rbrace} \Bh_{5} \Bh_{1}\herm \Bw_{\lbrace 135 \rbrace}  \\
        & \Bb_{ \lbrace 45 \rbrace} [j]=  \Ba_{\lbrace 145 \rbrace} [j] \Bv_{\lbrace 145 \rbrace} \Bh_{5} \Bh_{1}\herm \Bw_{\lbrace 145 \rbrace}.  
    \end{aligned}
\end{equation}
 
 Here, since $ j \in [N_{\rm T}]$, we have a system of equations with $N_{\rm T}$ equations and $N_{\rm T}$ unknowns. Therefore, by following the same approach as in~\eqref{eq: y hat def}-\eqref{eq: f k Q j A}, user $1$ can recover $A_{23}$, $A_{24}$, $A_{25}$, $A_{34}$, $A_{35}$, and $A_{45}$. Accordingly, these signals are decoded to the set of subpackets, expressed as:
 \begin{equation}
     \begin{aligned}
        \left\lbrace  \hat{W}^{1}_{\CQ \backslash \lbrace 1 \rbrace } : \CQ \subset [5], \left\vert \CQ \right\vert =3, 1 \in \CQ \right\rbrace.
     \end{aligned}
 \end{equation}

 Consequently, user $1$ retrieves $6$ subpackets of its requested file during the DL step, while it has stored $\gamma \binom{K}{t} \binom{K-t-1}{L-1}=4$ subpackets of its requested file before starting the UL-DL steps, and hence, it can decode its requested file $W^{1}$.

For the achievable NDT of the DL step, there are $N_{\rm DL}=N_{\rm S}N_{\rm T} = 6$ transmissions, and  $Q = \frac{1}{\binom{K}{t} \binom{K-t-1}{L-1}}=\frac{1}{10}$. As a result, by using~\eqref{eq: T DL def.} and Theorem~\ref{theorem: DL}, the NDT for the DL step is obtained as
 \begin{equation}
       \begin{aligned}
           T_{\rm DL} = \frac{N_{\rm DL}}{\binom{K}{t} \binom{K-t-1}{L-1}} = \frac{K-t}{t+L}=\frac{3}{5}.
       \end{aligned}
   \end{equation}
\end{document}